\documentclass[sigconf]{acmart}
\usepackage{amsmath,amsfonts}
\usepackage{algorithmic}
\usepackage{graphicx}
\usepackage{textcomp}
\usepackage{xcolor}
\usepackage{xspace}
\usepackage{float}
\usepackage{multirow}
\usepackage{lipsum}
\usepackage{pifont}
\usepackage{subcaption}
\usepackage{numprint}
\usepackage{enumitem}
\newcommand{\cmark}{\ding{51}}%
\newcommand{\xmark}{\ding{55}}%
\newcommand{\name}{\textit{TickTock}\xspace}
\newcommand{\mic}{mic\xspace}
\newcommand{\Mic}{Mic\xspace}
\newcommand{\mics}{mics\xspace}
\newcommand{\Mics}{Mics\xspace}
\newcommand{\revision}[1]{{\color{black}#1}}
\newcommand{\fclk}{$f_{\textrm{mic}}$}
\newcommand{\lclk}{$l_{\textrm{mic}}$}
\newcommand{\lold}{$\mathcal{L}_{2012}$}
\newcommand{\lmid}{$\mathcal{L}_{2017}$}
\newcommand{\lnew}{$\mathcal{L}_{2021}$}
\acmConference[CCS '22]{Proceedings of the 2022 ACM SIGSAC Conference on Computer and Communications Security}{November 7--11, 2022}{Los Angeles, CA, USA}
\acmDOI{}
\renewcommand\footnotetextcopyrightpermission[1]{} 
\begin{document}
\title{{\em TickTock}: Detecting \revision{Microphone Status in Laptops} Leveraging Electromagnetic Leakage of Clock Signals}
\subtitle{(Extended Version)}\subtitlenote{This is an extended version of the conference paper in the proceedings of ACM CCS 2022 with the same title.}
\author{Soundarya Ramesh}
\email{sramesh@comp.nus.edu.sg}
\affiliation{
    \institution{National University of Singapore}
    \country{Singapore}
}
\author{Ghozali Suhariyanto Hadi}
\email{ghadi@comp.nus.edu.sg}
\affiliation{
    \institution{National University of Singapore}
    \country{Singapore}
}
\author{Sihun Yang}
\email{ddyshsh@yonsei.ac.kr}
\affiliation{
    \institution{Yonsei University}
    \country{Republic of Korea}
}
\author{Mun Choon Chan}
\email{chanmc@comp.nus.edu.sg}
\affiliation{
    \institution{National University of Singapore}
    \country{Singapore}
}
\author{Jun Han}
\email{junhan@cyphy-lab.org}
\affiliation{
    \institution{Yonsei University}
    \country{Republic of Korea}
}
\renewcommand{\shortauthors}{Ramesh et al.}

\begin{abstract}
We are witnessing a heightened surge in remote privacy attacks on laptop computers. These attacks often exploit malware to remotely gain access to webcams and microphones in order to spy on the victim users. While webcam attacks are somewhat defended with widely available commercial webcam privacy covers, 
unfortunately, there are no adequate solutions to thwart the attacks on \mics 
despite recent industry efforts. 
As a first step towards defending against such attacks on laptop \mics, we propose \name, a novel \mic \textit{on/off} status detection system. 
To achieve this, \name externally probes the electromagnetic (EM) emanations that stem from the connectors and cables of the laptop circuitry carrying \mic clock signals. This is possible because the \mic clock signals are only input during the \textit{\mic recording} state, causing resulting emanations.
We design and implement a proof-of-concept system to demonstrate \name's feasibility. Furthermore, we comprehensively evaluate \name on a total of 30 popular laptops executing a variety of applications to successfully detect \mic status in 27 laptops. \revision{Of these, \name consistently identifies \mic recording 
with high true positive and negative rates.}
\end{abstract}

\maketitle
\pagestyle{plain}

\section{Introduction}
\label{sec:introduction}
Remote privacy attacks on modern day laptops continue to cause significant social problems.  
For example, remote attackers inject malware to gain access to webcams to stealthily spy on victims by disabling the webcam's indicator LED~\cite{webcam-attack-1, webcam-attack-2, webcam-attack-3, brocker2014iseeyou}. To defend against such attacks, users often place commercially available webcam privacy covers to physically block the webcams~\cite{webcam-cover-1, webcam-attack-2}. 
Exacerbating the problem, there are also reported attacks that \textit{spy on laptop microphones} -- including zero-day vulnerabilities and stalker-installed malware that stealthily eavesdrop from victims' laptops~\cite{facetime-bug, messenger-bug,fruitfly-2,fruitfly}. Moreover, bugs have been identified until recently in popular video calling apps, such as Zoom, which captured audio on Mac OS, 
even after the meeting had ended~\cite{zoom-mic-bug}. Unlike webcam covers, there are no immediately adequate solutions to defend against 
\mic-based eavesdropping. 

To defend against such attacks, companies such as Purism are pushing forward new laptop designs with hardware kill switches for \mics, which can cut off power supply to the \mics when not in use~\cite{kill-switch-2, kill-switch}. Apple designed a hardware disconnect feature for Macbook 2019 and later models, which disables the \mic whenever the lid is closed~\cite{kill-switch-apple}. Dell has updated its drivers on newer devices to allow for disabling \mics at the operating systems level~\cite{dell-mic-disable}. Furthermore, several operating systems such as Windows 10 and Mac OS 12 are providing indicators on screen during \mic usage for increased user awareness~\cite{mac-indicator, windows-indicator}.

While these efforts are promising first steps, they all suffer from significant shortcomings. First, these solutions require users to trust the implementation of the laptop manufacturers or the operating systems, both of which have been compromised by attackers several times in the past~\cite{brocker2014iseeyou, rat-malware, dma-attacks-laptop} or that the manufacturers themselves could be malicious. Second, these solutions are incorporated in only a small fraction of devices, hence most current day laptops do not have a way to detect/prevent eavesdropping. 
%

\begin{figure}[!t]
    \centering
    \includegraphics[width=1\linewidth]{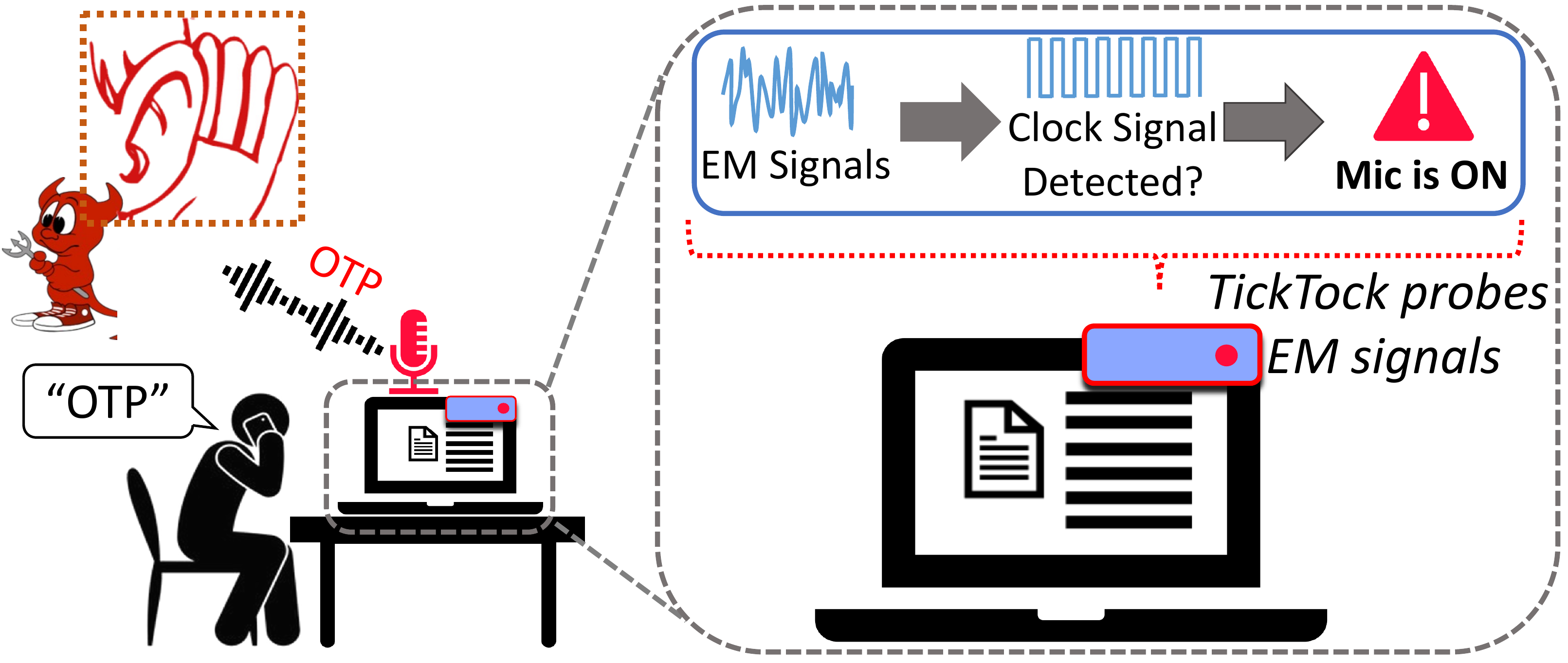}
    \caption{Figure depicts a scenario where a user places \name device (that equips an electromagnetic (EM) probe) in close vicinity of his/her laptop in order to detect a possible \mic-based eavesdropping attack, namely by determining if its \mic is \textit{ON} or \textit{OFF}. \name is able to do so based on the presence/absence of EM emanation of clock signals that are input to the \mic in the laptop circuitry.} 
    \label{fig:intro-fig}
\end{figure}
\begin{figure}[!t]
        \centering
        \includegraphics[width=0.9\linewidth]{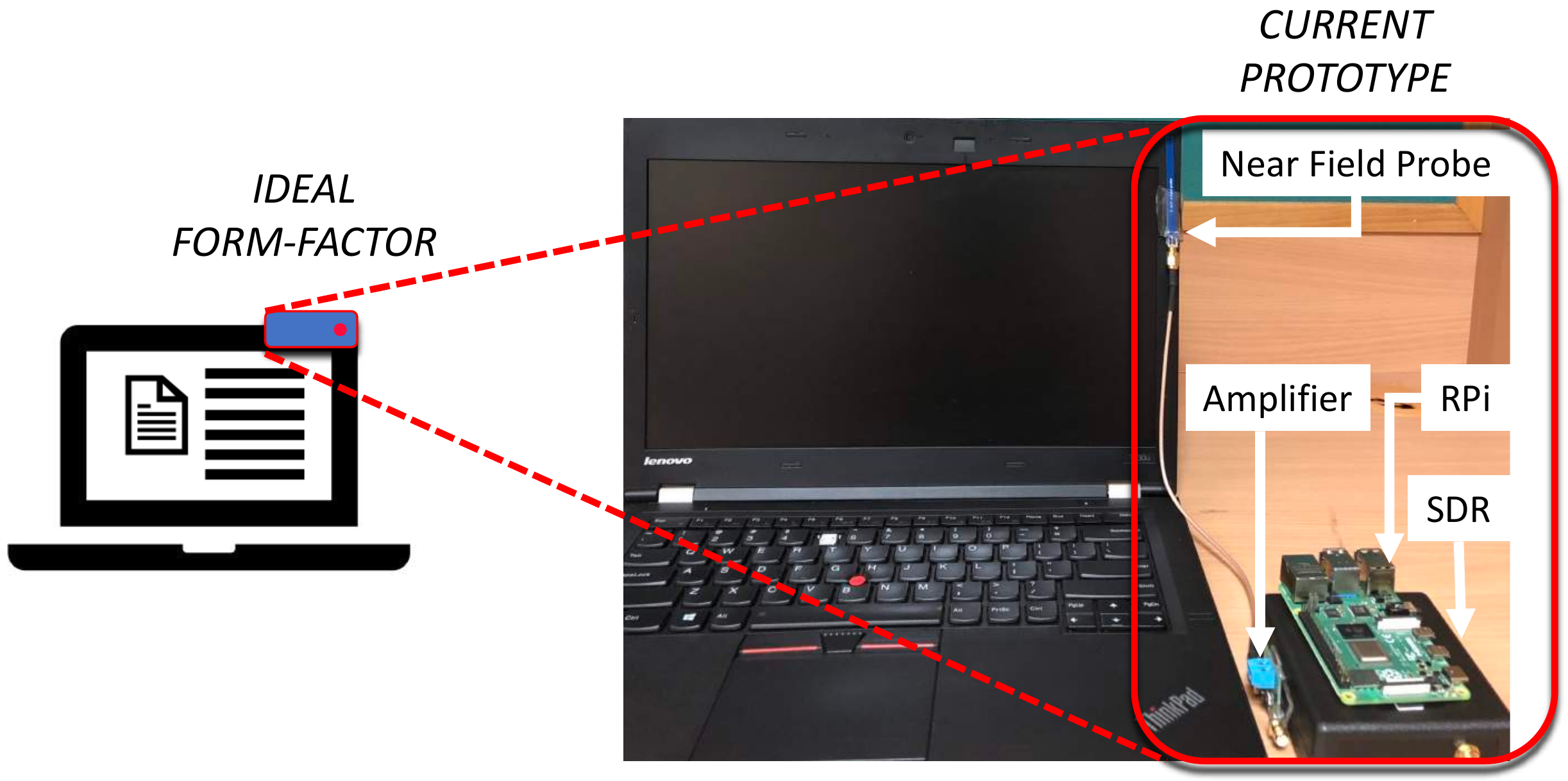}
        \caption{Figure depicts fully functioning prototype of \name, consisting of different components stacked to the side of the laptop. However, as depicted on the left, we envision \name with a form-factor similar to a small USB drive to be placed in contact with the laptop's exterior housing.}
        \label{fig:prototype}
\end{figure}

%
The aforementioned shortcomings lead us to the following research question: 
\textit{Can we design a novel \mic-based eavesdropping attack detection system that -- 
(1) is robust to powerful remote attackers,  (2) is applicable to existing laptops without any modifications, and (3) places limited trust on device manufacturers?} 
To this end, we propose \name that utilizes the phenomenon that digital MEMS \mics equipped in commodity laptops, when turned on (i.e., while recording), \textbf{\textit{emanate electromagnetic (EM) signals}}. 
The emanation stems from the cables and connectors that carry the clock signals to the \mic hardware, 
ultimately to operate its analog-to-digital converter (ADC) (see \S\ref{sec:bg-mic}). \name captures this leakage to \textit{identify the on/off status} of the laptop \mic. 
Figure~\ref{fig:intro-fig} depicts the process of utilizing \name. The user locates \name device -- consisting of a small EM probe --- on the external housing of the laptop near the leakage location. When the \mic starts recording, \name detects the clock signal and alerts the user (e.g., LED lights up). We envision \name to have a form-factor, similar to a USB drive (Figure~\ref{fig:intro-fig}), that can be adhered to the external of the laptop for detecting \mic \textit{on/off} status. However, \name's current fully-functional prototype has a table-top form-factor (Figure~\ref{fig:prototype}), but we see several opportunities to miniaturize this further (see \S\ref{sec:discussion}).  

Designing \name leads to three significant challenges. First, the \textit{frequency} of the \mic clock signal is unknown as its value varies across devices (typically ranging between 512 kHz to 4.8 MHz), particularly depending on the audio codec chip. Second, the \textit{location} of maximum leakage of the EM signals due to the \mic clock signals is also unknown, as it depends on the underlying location of the leaking cables and connectors. Third, as the EM signals captured typically include noise from neighbouring signal lines, we need to devise a robust mechanism for preventing false predictions. 

To overcome the aforementioned challenges, \name uses a one-time bootstrapping process per device model to infer the \mic clock frequency (\fclk), as well as the maximum leakage location (\lclk). In order to solve the third challenge of robust detection of clock signals in the presence of noise, \name leverages both the fundamental clock frequency as well as the \textit{harmonics}, which are multiples of the fundamental frequency, to improve detection accuracy.

\name has several advantages. First, adversaries with software capabilities cannot evade our detection as \name's approach relies on EM leakage due to the \mic hardware, hence making it robust against powerful remote attackers. Second, as \name's detection system is completely external to the devices themselves, it places minimal trust on the device manufacturers and software vendors.

We evaluate \name on a total of 30 laptops, with EM signals collected for over ten hours to demonstrate that \name \revision{\xspace detects \mic activities across most laptop brands we tested} including Lenovo, Dell, HP and Asus. We comprehensively evaluate \name's performance over different \mic-based applications (e.g., Zoom, Audacity), non-\mic based applications (e.g., Google News, YouTube), as well as different audio driver implementations (e.g., Ubuntu vs. Windows). In addition, we also evaluate its real-time performance, as well as its robustness to 
EM noise. From our analysis, \name successfully identifies the \mic clock frequency 
in 27 out of 30 tested laptops. \revision{Of the 27 laptops, \name consistently predicts \mic activities with high true positive and true negative rates. }

\section{System and Threat Model}
\label{sec:model}
\revision{
We present the system and threat models of \name. 

\noindent\textbf{\textit{System Model.}}
The \textit{goal} of \name is to identify \mic recording status (i.e., \textit{on/off}) in victim-owned devices, such as his/her laptop. We define a \mic to be \textit{recording} (i.e., \textit{\mic on}), whenever it captures physical acoustic signals and converts them into digital signals. Hence, we do not distinguish between cases where the digital signals from the \mic are saved to memory vs. when they are discarded, and consider both as \textit{recording}. 
\name is \textit{constrained} to only capture EM leakages from close contact on a device (e.g., from external housing of a laptop). Hence, we do not consider \mic status detection in spying devices (e.g., audio bugs hidden in a room). 
Furthermore, \name is \textit{constrained} to only detect \mic status in devices with \textit{digital} \mics (i.e., \mics that require clock signals for their operation). }

\noindent\textbf{\textit{Threat Model.}} 
In designing \name, we consider an attacker with the following goal and capabilities. 
The attacker's \textit{goal} is to stealthily capture audio from the \mic of the victim's laptop. The attacker's \textit{capabilities} include launching \textit{remote} attacks with unconstrained software capabilities. Specifically, we consider powerful attackers that may control malicious or compromised applications, and are capable of exploiting kernel vulnerabilities to modify the audio drivers. However, we assume that the attacker does not have physical access to the laptop, and cannot modify the hardware (e.g., embed a standalone audio bug within the laptop). 

\section{\name Usage Scenarios}
\label{sec:usage-scenarios}
\begin{figure}[!t]
    \centering
    \includegraphics[width=0.9\linewidth]{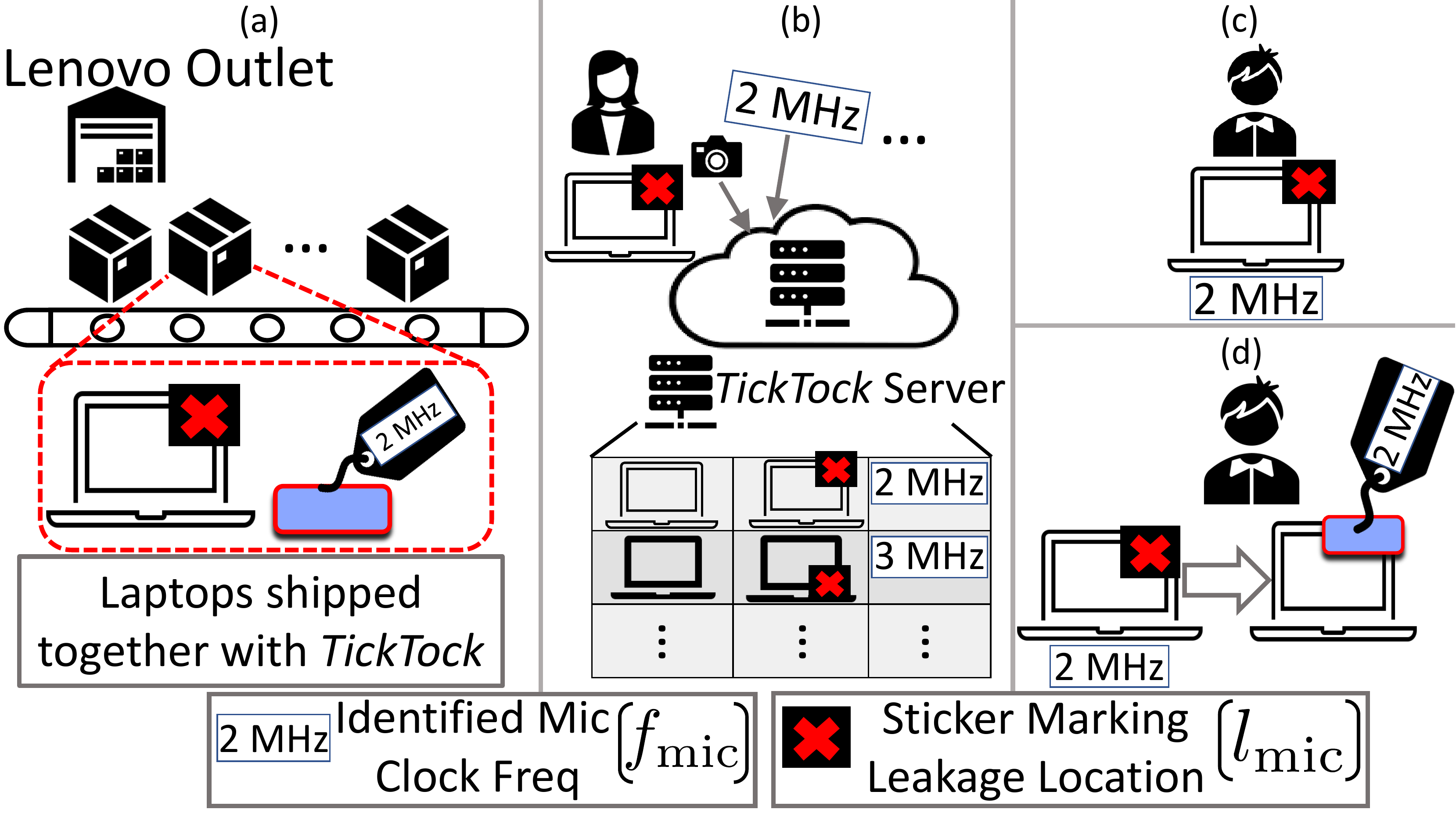}
    \caption{\textit{Bootstrapping} scenarios -- (a) depicts bootstrapping performed by laptop manufacturers, e.g., Lenovo, who subsequently ship laptops with stickers denoting leakage location, \lclk, and an accompanying \name device set to detect \mic clock frequency, \fclk. (b) depicts a crowd-sourcing scenario where users upload \fclk, along with an image of \lclk~ to \name's public server. (c) depicts a scenario where a user locally performs bootstrapping. (d) depicts a deployment scenario, where a user deploys \name to detect \mic \textit{on/off} status by placing the \name device at the location of the sticker, and by setting the \name device to detect \fclk.}
    \label{fig:motivating-scenario}
\end{figure}

This section presents the potential usage scenarios of \name. 

\noindent\textbf{Bootstrapping Scenario.} \name requires a \textit{one-time bootstrapping phase}, to infer \textit{\mic clock frequency}, \fclk, and \textit{maximum EM leakage location}, \lclk, for each device model. We present three scenarios we envision for different entities performing bootstrapping. 

\noindent\textit{(1) Bootstrapping by Manufacturers.} 
Laptop manufacturers, e.g., Lenovo, can perform the bootstrapping phase for their products. Following which, as depicted in Figure~\ref{fig:motivating-scenario}(a), they ship each of their laptops with -- (a) an accompanying \name device, that is set to detect the \fclk~(e.g., 2 MHz), and (b) a sticker placed on the laptops (e.g., in the top-right corner), in order to mark \lclk. 

\noindent\textit{(2) Crowd-sourced Bootstrapping.} 
A crowd-sourced approach (see Figure~\ref{fig:motivating-scenario}(b)) is where average users conduct bootstrapping on one/more devices, and upload detected \fclk~ and \lclk~ to \name's server. This information can be utilized when users deploy \name. 

\noindent\textit{(3) User-level Bootstrapping.} 
\name's bootstrapping is conducted by the user (Figure~\ref{fig:motivating-scenario}(c)) intending to use \name on his/her laptop.

\noindent\textbf{Deployment Scenario.} 
To use \name (Figure~\ref{fig:motivating-scenario}(d)), the user leverages the bootstrapping information, 
and sets \fclk~ on the \name device. Subsequently, the user places the \name device on 
\lclk~ to enable \name to function as a \mic \textit{on/off} status indicator.   


\section{Background}
\label{sec:background}
We provide background on the role of clock signals in determining \mic status, 
why they leak, and how their leakage can be detected.  

\begin{figure}[!t]
    \centering
    \includegraphics[width=0.9\linewidth]{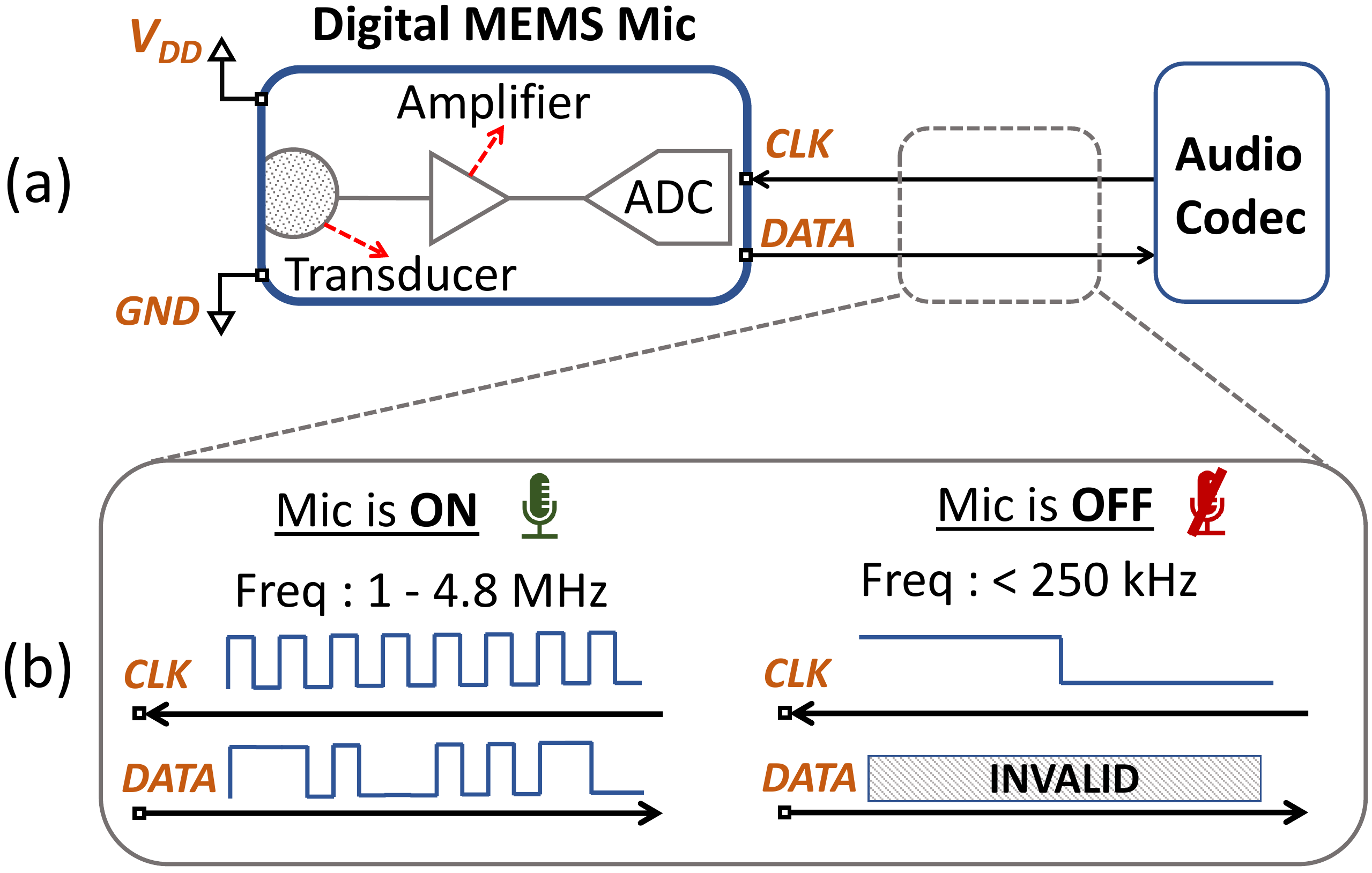}
    \caption{Figure (a) depicts the functioning of a digital MEMS \mic, which takes as input a clock signal, in order to digitize electrical signals, and ~\revision{(b) depicts the difference in \mic clock frequency when the \mic is \textit{on} vs. when it is \textit{off}.}}
    \label{fig:bg-mics}
\end{figure}
%
\subsection{Digital MEMS \Mics}
\label{sec:bg-mic}
Laptops typically contain Micro-Electro-Mechanical systems (MEMS) \mics mainly due to their compact form-factor and better noise performance~\cite{mems-mic-1, mems-mic-2}. \revision{Amongst them,  
\textit{digital} MEMS mics, which are immune to electromagnetic interference (EMI), are a preferred alternative. This is because in laptops, the long cables or PCB traces carrying \mic data lines 
may run close to electromagnetic disturbances such as the laptop's liquid crystal display~\cite{mems-mic-3, mems-mic-4}}. 
Digital \mics sample the analog signal to output data in the form of discrete, high amplitude signals, alternating between the two extreme voltage levels -- representing 0 and 1 respectively. As depicted in Figure~\ref{fig:bg-mics}(a), digital \mics contain an analog-to-digital converter (ADC) within the \mic housing, and the ADC's operation is driven by an input clock signal. 
Furthermore, these \mics support a wide range of operating clock frequencies from about 512 kHz to 4.8 MHz~\cite{knowles-clock,invensense-clock}. 

%
\subsubsection*{\textbf{\revision{Role of Clock Signals in Mics}}}
\label{sec:bg-clock-role}

~\revision{In digital MEMS ~\mics, clock signals function as a \textit{control signal} that can switch the \mic between several power modes. 
As depicted in Figure~\ref{fig:bg-mics}(b), when the \mic is provided with a clock signal in the frequency range around $1-4.8$ MHz, it enters \textit{active mode} where it consumes about 0.5 mA of current, and hence is capable of capturing audio ~\cite{knowles-i2s-1, knowles-pdm-1, knowles-pdm-2}. On the other hand, when the \mic is provided with clock signals whose frequencies are below 250 kHz, 
the \mic enters \textit{sleep mode}~\footnote{~\revision{Some \mics also support a \textit{low-power mode} for clock frequencies from 512 kHz to 1 MHz, suitable for wake-word detection in voice-enabled applications.}} in order to reduce power consumption ($\approx 40$ $\mu A$)~\cite{infineon-mic}. 
In this work, we identify this difference in clock frequency when the \mic is in active mode (i.e., the \mic is \textit{on}), and when in sleep mode (i.e., \mic is \textit{off}), from the EM leakage signals, in order to infer \mic's \textit{on/off} status.}

\begin{figure}[!t]
    \centering
    \includegraphics[width=0.75\linewidth]{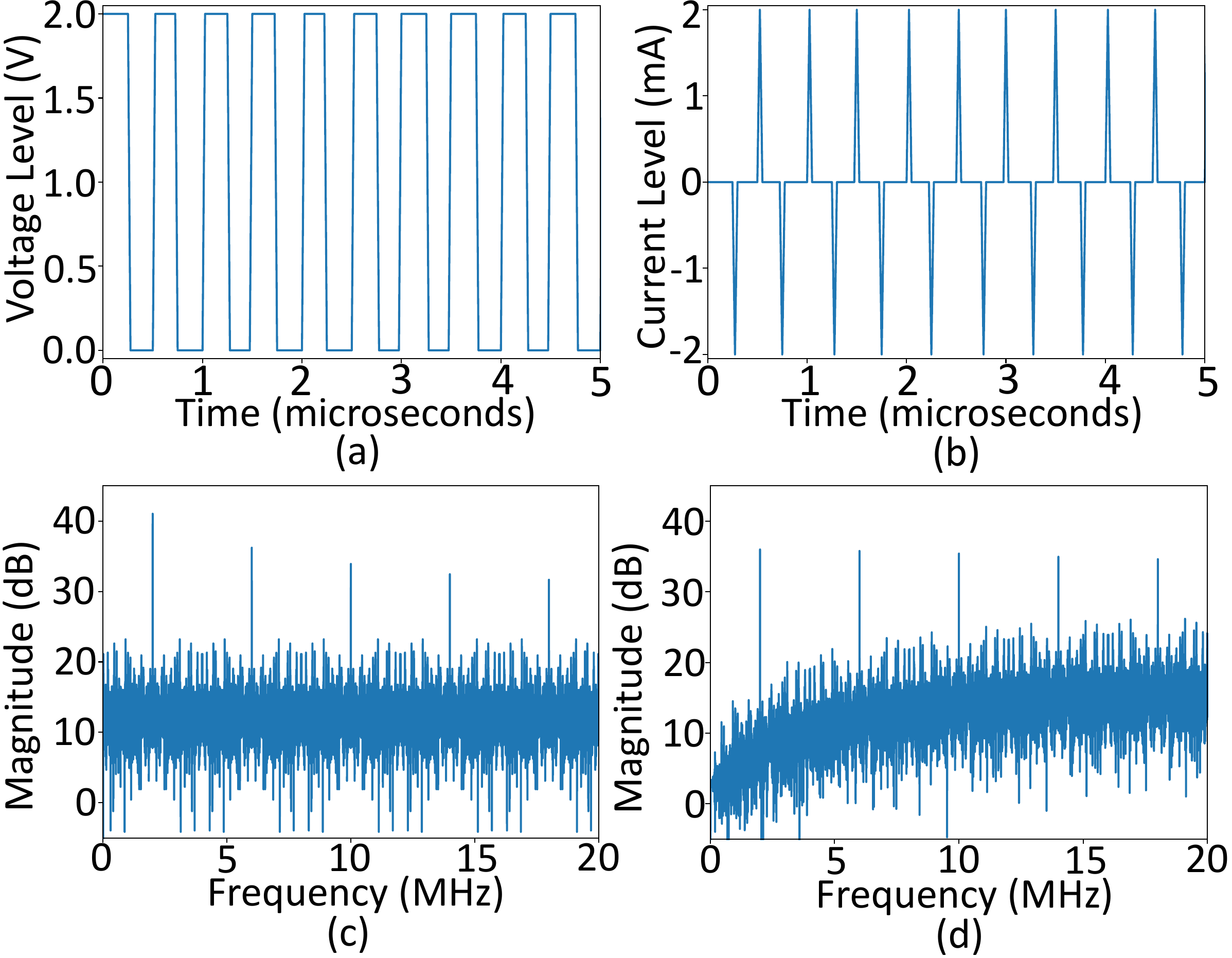}
    \caption{Figure depicts the \revision{ideal representation of clock signals} in time-domain as (a) voltage and (b) current trends; in frequency-domain as (c) voltage, and (d) current trends.} 
    \label{fig:bg-clock}
\end{figure}
%
\subsection{Clock Signals and their Detection}
\label{sec:bg-clock}

Clock signals, expressed as voltage in the time domain (Figure~\ref{fig:bg-clock}(a)), are periodic square waves with a fixed time period (denoted by T), and has a fundamental frequency, $f_{clk} = \frac{1}{T}$. When observed as current in the time  domain (Figure~\ref{fig:bg-clock}(b)), clock signals are seen as a series of impulses, as the current flows only during a voltage change. Nevertheless, this signal has the same time period, T, and fundamental frequency, $f_{clk}$.

Due to their periodic nature, as well as the short rise-time for transition between voltage levels (sub-microseconds), clock signals 
concentrate their energy in the fundamental clock frequency, $f_{clk}$, as well as its odd harmonics (i.e., $3\small{\times}, 5\small{\times}, 7\small{\times}, \dots$) (see Figures~\ref{fig:bg-clock}(c) and ~\ref{fig:bg-clock}(d)). 
Furthermore, if the clock signals spend unequal time in the high and low voltage states
(i.e., if the clock duty cycle deviates from 50\%), the radiated signal will additionally include even harmonics, i.e., $2\small{\times}, 4\small{\times}, 6\small{\times}$, and so on, of the fundamental clock frequency, $f_{clk}$. 

%
\begin{figure}[!t]
        \centering
        \includegraphics[width=0.75\linewidth]{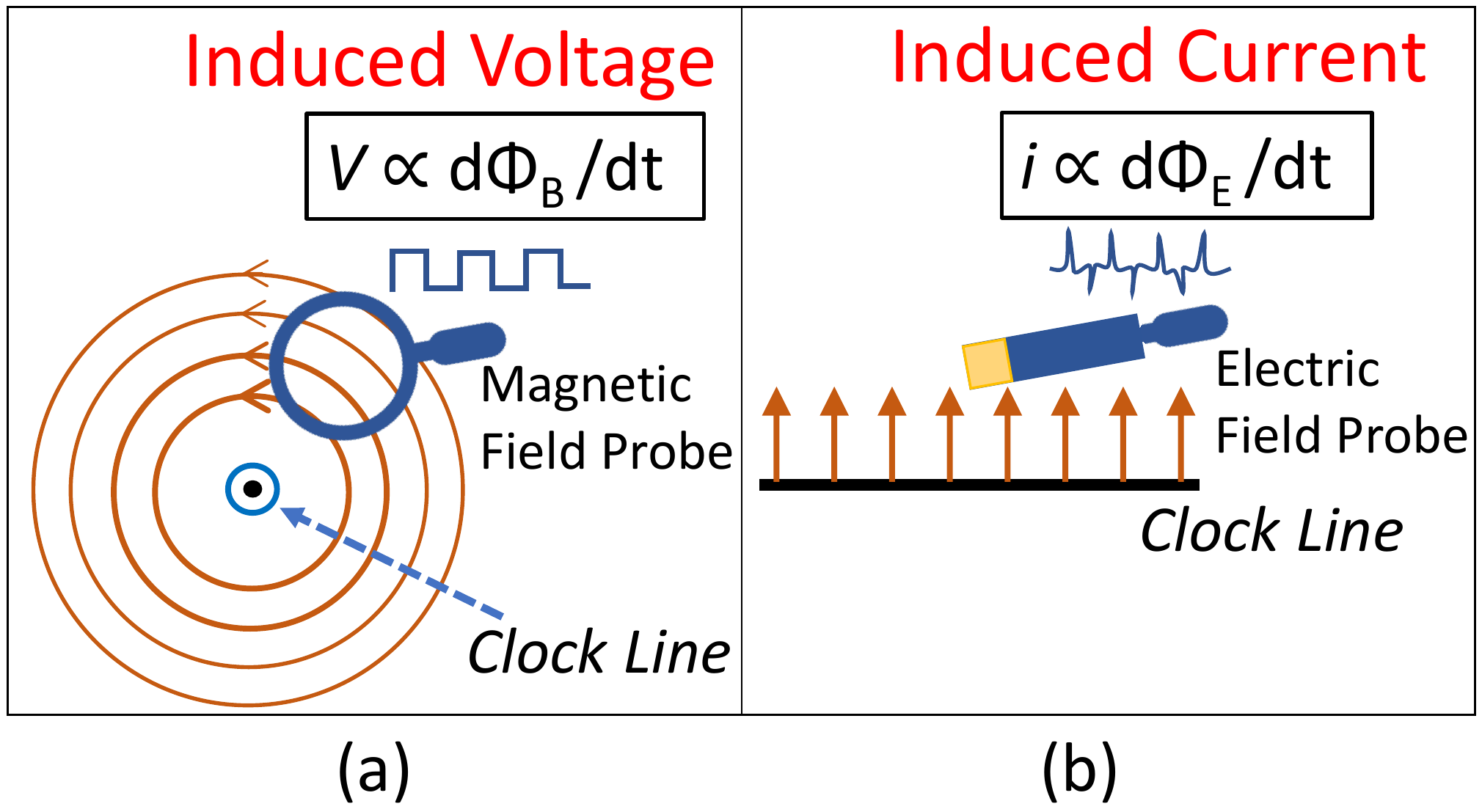}
        \caption{Figure depicts the working of the (a) magnetic field (H-field) probe, and (b) electric field (E-field) probes.} 
        \label{fig:bg-probesv2}
        \vspace{-0.5cm}
\end{figure}
%

\begin{figure*}[!t]
        \centering
        \includegraphics[width=0.9\linewidth]{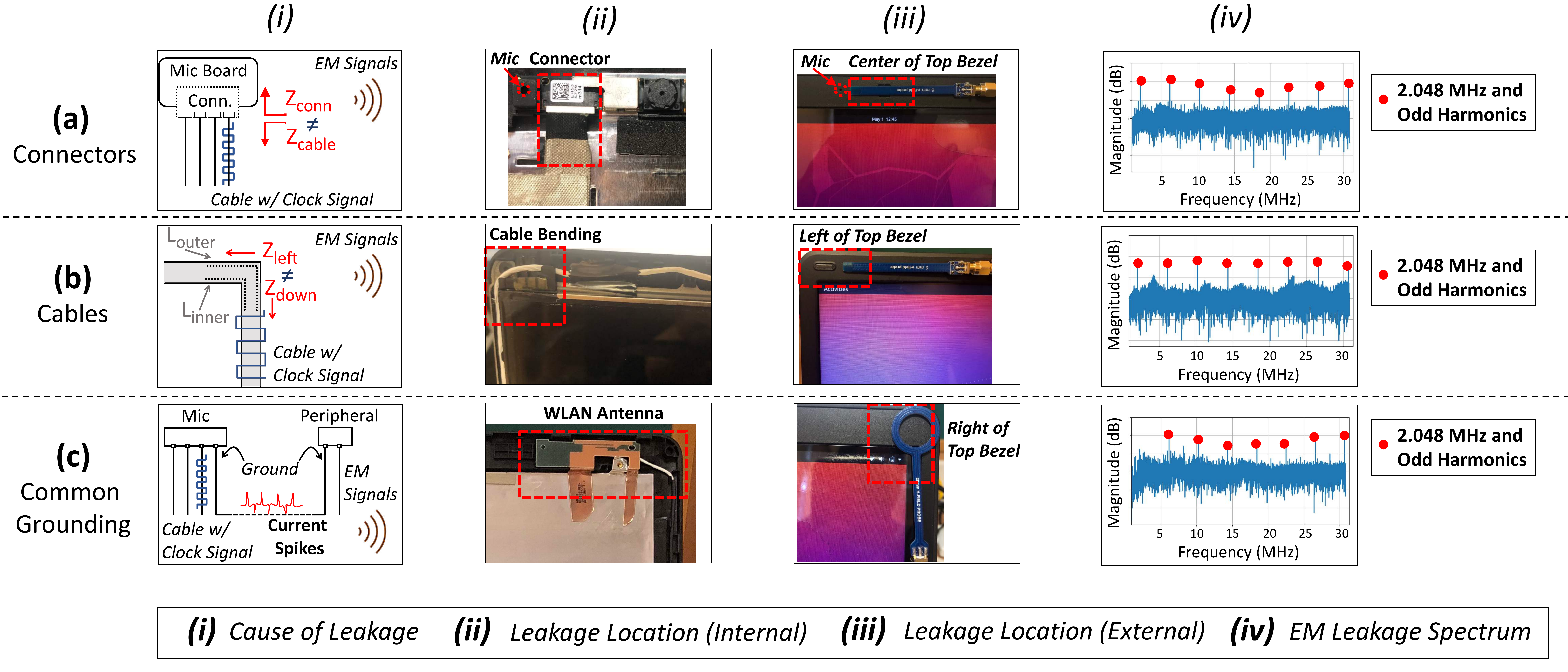}
        \caption{~\revision{Figure depicts the cause of leakage, leakage location (internal and external) as well as the captured EM leakage signals (with the \mic \textit{on}) due to three different leakage factors, namely (a) connectors, (b) cables, and (c) common grounding.}}
        \label{fig:feasibility-teardown}
\end{figure*}

\subsubsection{\textbf{Detecting Clock Signals}} 
\label{sec:bg-detect}
We utilize near-field probes, namely –- magnetic field (or H-field) and electric field (or E-field) probes, to capture the variations in the magnetic field and electric field due to clock signals respectively.

Electromagnetic (EM) signals exhibit different behavior in the near-field (i.e., region within about one wavelength of the leaked signal frequency) as compared to the far-field (those beyond one wavelength). In the near-field, the electric and magnetic fields exist independently, where one could dominate the other depending on the source of clock signal leakage. In the far-field, the two fields are coupled together to form the EM field. 
As EM signals significantly attenuate before reaching the far-field, especially for lower frequencies (in MHz), we perform our detection in the near-field. 

A simplified view of how the probes work is depicted in Figure~\ref{fig:bg-probesv2}. The magnetic-field probe outputs a voltage proportional to the rate of change of magnetic flux passing through the loop (Figure~\ref{fig:bg-probesv2}(a)), while the electric-field probe has an induced current proportional to the change in electric-field experienced by the conductor at the probe's tip (Figure~\ref{fig:bg-probesv2}(b)). In particular, for the magnetic-field probe, the loop size determines its sensitivity to weak EM signals, where a loop with larger radius is more sensitive.

\subsection{Factors of Clock Signal Leakage}
\label{sec:bg-factors}
\revision{As depicted in Figure~\ref{fig:feasibility-teardown}, we identify three potential factors, namely -- (a) connectors, (b) cables, and (c) common grounding, that lead to EM leakage of clock signals. For each factor, we explain their causes of leakage based on theory (part (i) in figure), and identify the exact leakage location in a laptop that leaks due to the factor discussed (part (ii)). Subsequently, we capture EM traces from the exterior of the laptop (part (iii)), to finally obtain the EM leakage spectrum containing the ~\mic clock signals (part (iv)). 
In order to capture the leaked EM signal, we place the near-field probe at the leakage location, and utilize the setup described in Section~\ref{sec:feasibility-setup}. 
We now explain each leakage factor in detail below. }

\revision{\subsubsection{\textbf{Leakage (a) -- Connectors}} 
\label{sec:bg-leakage-1}
Impedance mismatch in connectors 
is a major contributor to EM emanation. 
As depicted in Figure~\ref{fig:feasibility-teardown}(a), when the impedance values of two adjacent elements, e.g., a connector ($Z_{\textrm{conn}}$) and a cable carrying clock signals ($Z_{\textrm{cable}}$) are mismatched, part of the transmitted signal can be reflected and emitted as EM signals. The amount of reflection, or \textit{reflection ratio}, can be approximated as: $\frac{(Z_2-Z_1)}{(Z_2+Z_1)}$, where $Z_1$ and $Z_2$ refer to the impedance of the source element and the receiving element, respectively. This reflection ratio is directly proportional to the amount of EM emission. 
Such EM emission issues occur when circuit designers do not take into account the additional impedance that may be produced on cables while carrying high frequency signals. 

In order to confirm the theory, we perform a teardown on Dell Latitude E5570 laptop where the connector is adjacent to the \mic. 
As depicted in Figure~\ref{fig:feasibility-teardown}(a)-(ii), we identify the connector's location on the right side of the \mic. Furthermore, by placing an E-field probe on the laptop's exterior (Figure~\ref{fig:feasibility-teardown}(a)-(iii)) at the same location, we obtain the EM spectrum with the clock frequency (2.048 MHz) and its harmonics as depicted in the figure (Figure~\ref{fig:feasibility-teardown}(a)-(iv)), confirming that connectors indeed lead to EM leakage.}
%

\subsubsection{\textbf{Leakage (b) --  Cables}} 
\label{sec:bg-leakage-2}
\revision{As depicted in Figure~\ref{fig:feasibility-teardown}(b)-(i), sharp turns in cables and PCB traces change the impedance characteristic of the cables due to difference in the propagation delay resulting from unequal lengths between inner (i.e., $L_{\textrm{inner}}$) and outer sides (i.e., $L_{\textrm{outer}}$) of the PCB traces and cables.  
Consequently, these unaccounted impedance changes cause impedance mismatch between two sides of the cable (e.g., $Z_{\textrm{left}}$, and $Z_{\textrm{down}}$ as depicted in the figure), leading to EM emissions. We confirm this source of leakage by performing a teardown of a Fujitsu Lifebook\footnote{\revision{Note that we utilize different laptops to demonstrate different leakage factors that may dominate in different laptops.}} in which the microphone cables bend along the top-left corner of the laptop (Figure~\ref{fig:feasibility-teardown}(b)-(ii)). 
We identify the clock frequencies and their harmonics by placing the near-filed probe on the laptop's exterior at the same location (Figure~\ref{fig:feasibility-teardown}(b)-(iii),(iv)). }

Similar to bending of cables, usage of flexible PCBs (or flex cables) for connecting \mic board to the 
audio codec, can result in EM signal leakage due to their flexible nature. While adding grounding copper layers can shield flex PCBs from leakage, such additional makes the PCB rigid, hence ruining their utility~\cite{flex-cable-emi}. 

\subsubsection{\textbf{Leakage (c) -- Common Grounding}} 
\label{sec:bg-leakage-3}
\revision{As clock signals have high current slew rate (i.e., high $di/dt$), current spikes in \mic ground lines 
lead to similar spikes in 
other peripherals with shared ground (Figure~\ref{fig:feasibility-teardown}(c)-(i))~\cite{EMI-emission, ground-noise-murata}. 
Consequently, this results in EM emissions of \mic clock frequencies at locations distant from the \mic clock lines. 
In particular, we observe this phenomenon at the location of the WLAN antenna in the top-right corner of the bezel of Dell Latitude E5570 laptop as depicted in Figure~\ref{fig:feasibility-teardown}(c)-(ii).}

\section{Feasibility Study}
\label{sec:feasibility}
By means of preliminary experiments, 
we 
demonstrate the feasibility of \mic clock leakage signals serving as a proxy for \mic status. 

\subsection{Feasibility Setup}
\label{sec:feasibility-setup}
Our setup (Figure~\ref{fig:prototype}) consists of -- a test device (e.g., laptop), a near-field probe (E-field / H-field) that captures EM leakage signals, connected to a 27 dB wideband RF low-noise amplifier (with an input voltage of 10V DC) to amplify the weak EM signals, which is in-turn connected to an SDRPlay RSP-1A software defined radio (SDR) that captures and digitizes the signals in the frequency range of interest, and finally an RPi 4B, executing GNU Radio Companion software, that 
performs signal processing~\cite{gnuradio,sdrplay,rf-lna-minicircuits,near-field-probe, rf-lna-aliexpress}. 

\subsection{Clock Signals as Mic Status Indicators}
\label{feas:mic-activity}

\begin{figure}[!t]
    \centering
    \includegraphics[width=0.9\linewidth]{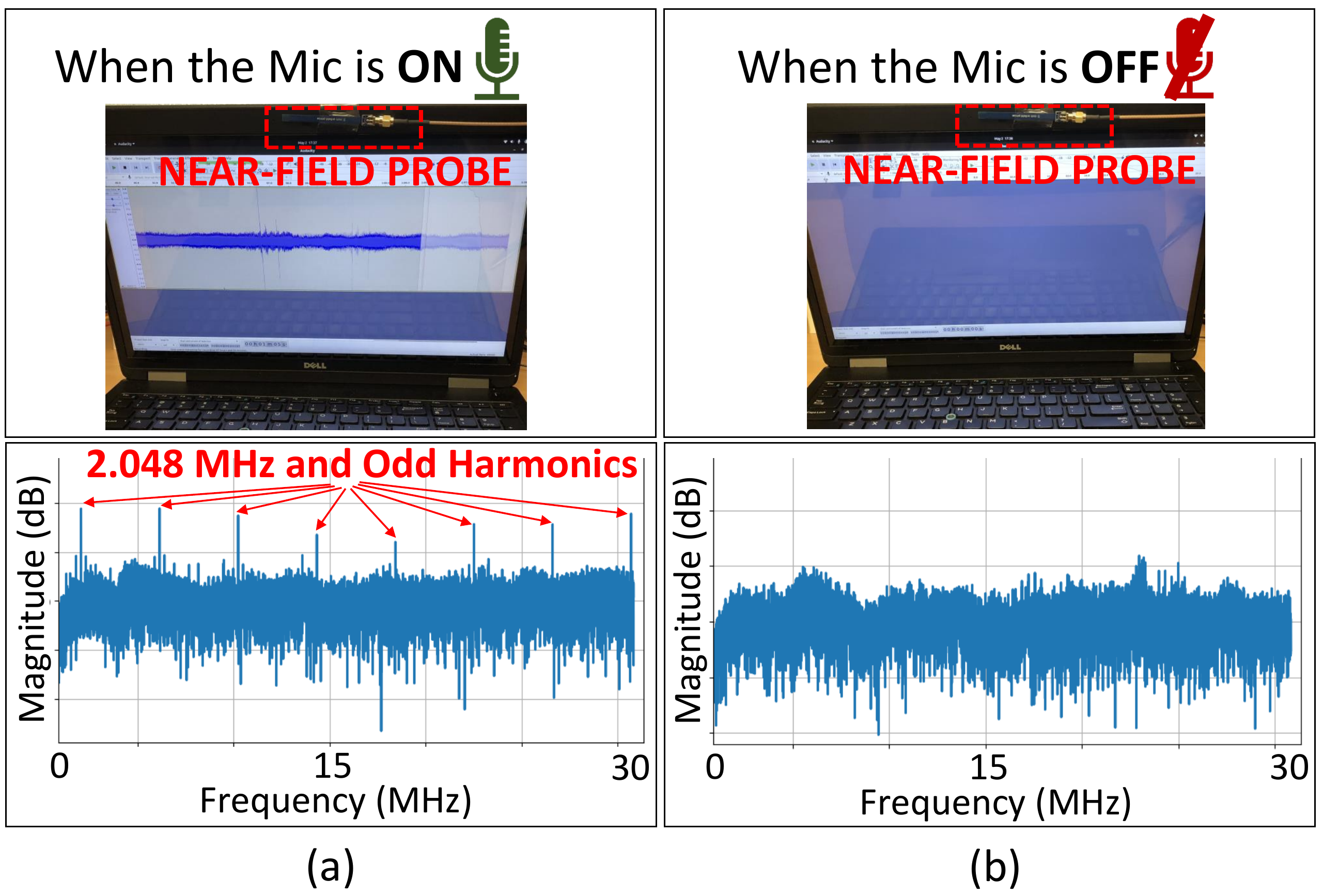}
    \caption{Figure depicts the EM leakage spectrum when -- (a) the \mic is \textit{on} (i.e., recording on \textit{Audacity} app), and (b) when the \mic is \textit{off}. The clock frequency (2.048 MHz) and harmonics are present only when the \mic is \textit{on}, hence indicating the feasibility of using EM signals to detect \mic \textit{on/off} status.
    }
    \label{fig:feasibility-spectrum}
    \vspace{-0.5cm}
\end{figure}

To confirm if the clock leakage signals can serve as a \mic status indicator, we perform experiments on Dell Latitude E5570 laptop. 
We place the near-field probe (specifically, E-field) at a location of maximum leakage, i.e., near the connector for this laptop (from Section~\ref{sec:bg-leakage-1}). 
As depicted in Figure~\ref{fig:feasibility-spectrum}, the \mic clock frequencies, i.e., 2.048 MHz and odd harmonics, are present \textit{only} when the \mic is \textit{on} (Figure~\ref{fig:feasibility-spectrum}(a)), and are absent otherwise (Figure~\ref{fig:feasibility-spectrum}(b)). This preliminary experiment suggests that presence/absence of clock signals can serve as proxy for indicating \mic's \textit{on/off} status respectively. 
In the following sections, we elaborate on \textit{how} we identify the \mic clock frequency and leakage location, as well as perform comprehensive experiments to test the robustness of \name.  

\section{Design and Implementation}
\label{sec:design}
We now present the design and implementation of \name. 

\begin{figure}[!t]
        \centering
        \includegraphics[width=0.9\linewidth]{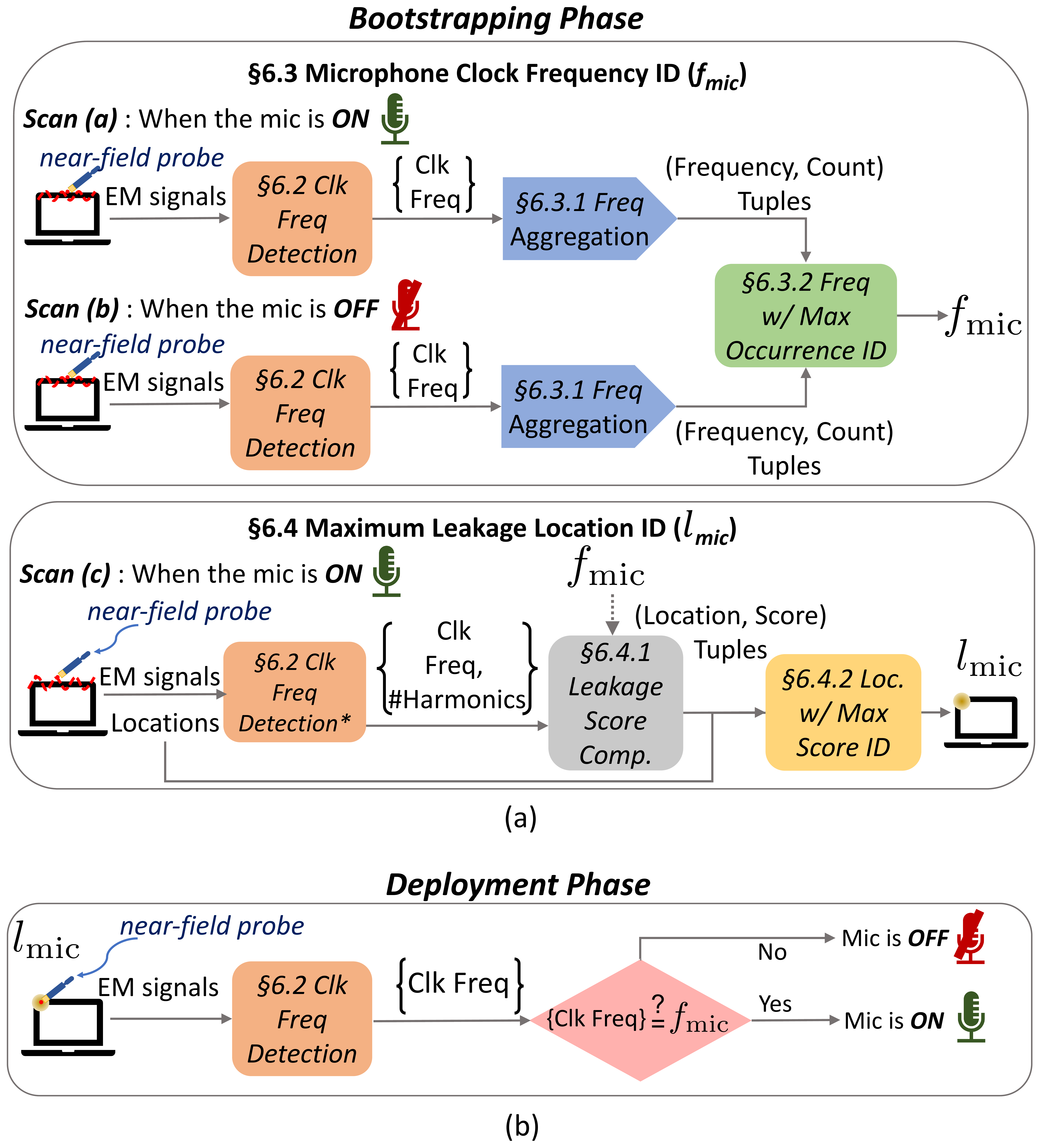}
        \caption{Figure depicts the system overview of \name. (a) depicts \textit{bootstrapping} phase where (1) we identify the \mic clock frequency, \fclk, as the frequency with maximum number of occurrences (or count) amongst the clock frequencies that only occur when the \mic is \textit{on}; 
        (2) we identify the \textit{maximum} leakage location,~\lclk, as the location where the \textit{leakage score} due the \mic clock signals 
        is maximum. 
        (b) depicts \textit{deployment} phase, where a user places the probe at \lclk, identified during bootstrapping, to detect \mic \textit{on/off} status of the device based on the presence/absence of its \fclk.}
        \label{fig:des-fig}
\end{figure}
%
\subsection{Design Overview}
\label{sec:des-overview}
\name leverages \mic clock signals from the leaked EM signals in order to serve 
as a \mic \textit{on/off} status indicator. Recall from \S\ref{sec:usage-scenarios} that ~\name's design consist of two phases, namely \textit{bootstrapping} and \textit{deployment} phases. \textit{Bootstrapping} is a one-time phase where we identify the \mic clock frequency, \fclk, as well as the \mic clock’s maximum leakage location, \lclk, of a certain device model. Subsequently in the \textit{deployment} phase, a user with the same device model, utilizes the identified frequency, \fclk, and location, \lclk, in order to predict \mic status of his/her device. 

During bootstrapping (Figure~\ref{fig:des-fig}(a)), we identify the \mic clock frequency, \fclk, by performing two scans (\textit{Scans (a) and (b)}), in the region near the \mic (e.g., laptop's top bezel) with a near-field probe -- once when the \mic is \textit{on}, and the second time when the \mic is \textit{off}. 
A \textit{scan} consists of probing multiple locations in a region (e.g., near the \mic) and observing the EM signals at each location over multiple time periods. 
Following the scans, we identify \fclk~ as the frequency that occurs uniquely \textit{only} in the EM signals when the \mic is \textit{on}, and has maximum 
occurrences compared to all other unique frequencies. Subsequently, in order to identify the maximum leakage location, ~\lclk, we perform a third scan (\textit{Scan (c)}), by determining a location with maximum \textit{leakage score}, which we compute based on the detection of the identified \fclk~ and its \textit{harmonics}.

In the \textit{deployment} phase (Figure~\ref{fig:des-fig}(b)), a user places the near-field probe at location, \lclk, identified during bootstrapping, for \mic status detection. 
\name predicts that the \mic is \textit{on}, \textit{only} if the set of detected clock frequencies from the EM signals (i.e., the output of \textit{Clock Frequency Detection} module in Figure~\ref{fig:des-fig}), contains \textit{exactly} one frequency which equals the \mic clock frequency, \fclk. Hence, \name predicts that the \mic is \textit{off} even when \fclk~ is detected along with other spurious frequencies, in order to minimize false predictions. However, \name tolerates some error margin (i.e., $\theta_{margin} = 10$ kHz) around \fclk, while predicting that the \mic is \textit{on}.

In the following subsections, we address the three main challenges of ~\name:   \textbf{\textit{Challenge 1:}} Clock Frequency Detection (\S\ref{sec:des-clk_freq_detection}), \textbf{\textit{Challenge 2:}} \Mic Clock Frequency Identification (\S\ref{sec:des-mic_freq_id}), and \textbf{\textit{Challenge 3:}} Maximum Leakage Location Identification (\S\ref{sec:des-probing_location_id}).

\begin{figure*}[!t]
        \centering
        \includegraphics[width=0.8\linewidth]{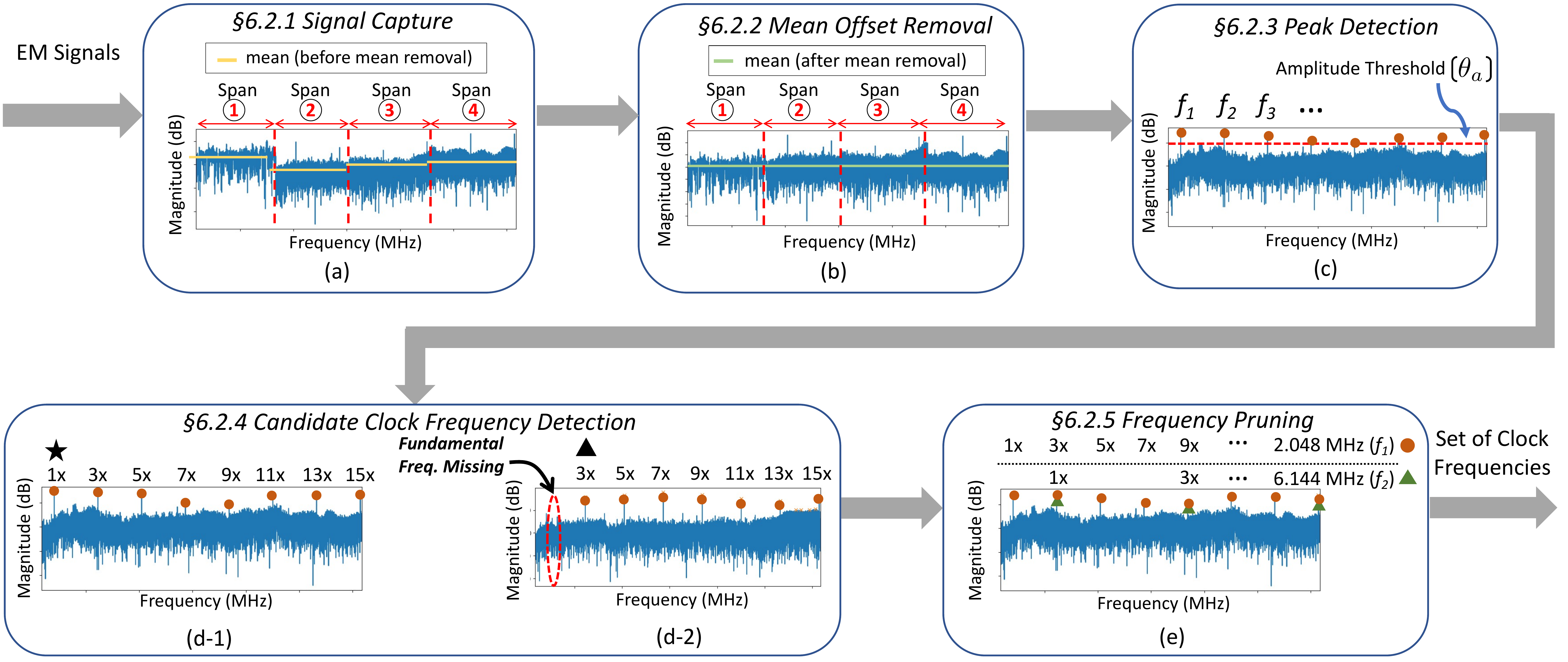}
        \caption{Figure depicts the overall design of \textit{Clock Frequency Detection} module. (a) depicts the integration of EM signals across four frequency spans (i.e., $n_s = 4$), to obtain the required larger bandwidth signal. (b) depicts mean removal performed to equalise noise floors across spans. (c) depicts peak detection where we identify frequency peaks (e.g., $f_1, f_2, f_3, \dots$), above a certain amplitude threshold, $\theta_a$. (d) depicts the identification of clock frequencies in the presence of harmonics. (e) depicts the pruning of clock frequencies, based on shared harmonics.} 
        \label{fig:des-clk-steps}
\end{figure*}
%
\subsection{Challenge 1: Clock Frequency Detection} 
\label{sec:des-clk_freq_detection}
One of the main challenges in robustly detecting clock frequencies is the presence of EM noise from neighbouring components or signal lines in the captured EM signals,  leading to detection of spurious frequencies. We overcome this issue by detecting their \textit{harmonics}, in addition to the fundamental frequency. Both \textit{Bootstrapping} and \textit{Deployment} phases utilize this module to take as input the EM leakage signals and output the set of detected clock frequencies.

Figure~\ref{fig:des-clk-steps} depicts the module overview. We capture EM signals, or \textit{traces}, across several frequency spans from the Software Defined Radio (SDR), and compute their spectrum (\S\ref{sec:des-clk-signal-capture}). Subsequently, we perform mean offset removal (\S\ref{sec:des-clk-mean-offset}), and identify the peaks in their frequency spectrum based on an amplitude threshold (\S\ref{sec:des-clk-peak-detection}). We leverage the detected peaks to identify a set of candidate clock frequencies based on the \textit{number of harmonics} detected as peaks (\S\ref{sec:des-clk-candidate-clk-id}). Finally, we input the candidate clock frequencies to a pruning stage and eliminate frequencies that are harmonically related to other more likely candidate clock frequencies (\S\ref{sec:des-clk-pruning}).

\subsubsection{Signal Capture}
\label{sec:des-clk-signal-capture}
In order to capture the EM trace from the SDR, we specify two parameters, namely -- center frequency ($f_c$), and bandwidth ($B$). By doing so, we obtain the EM trace information within frequency span, $[f_c-B/2, f_c+B/2]$. However, as the maximum bandwidth supported by many low-end SDRs may not be sufficient to detect the \mic clock frequency \textit{and} their first several harmonics, we sweep across several ($n_{s}$) adjacent frequency spans in order to obtain leakage signals with an overall larger bandwidth ($ = n_s B$). 
Subsequently, we compute the magnitude spectrum of each span (Figure~\ref{fig:des-clk-steps}(a)), and stitch all the spans together (here $n_s = 4$), to obtain the overall spectrum of the leaked EM signals.

\subsubsection{Mean Offset Removal}
\label{sec:des-clk-mean-offset}
We observe that the noise floors of different spans may be different due to different gain values across spans. This is a result of implementation of automatic gain control feature in several SDRs. We perform span-wise mean-offset removal in order to equalise the noise floors across spans (see Figure~\ref{fig:des-clk-steps}(b)).

\subsubsection{Peak Detection}
\label{sec:des-clk-peak-detection}
We first obtain the magnitude spectrum of the entire EM trace, we compute a set of peaks in the frequency domain, namely, $\mathbb{F}_{p} = \{f_1, f_2, f_3, \dots\}$ (see Figure~\ref{fig:des-clk-steps}(c)). The peaks satisfy a minimum amplitude cutoff, $\theta_{a}$, and are separated in frequency at least by a distance, $\theta_{d}$. The amplitude threshold, $\theta_{a}$, varies across devices, depending on the level of leakage of the \mic clock signals. However, the distance parameter, $\theta_{d}$, is fixed across all devices for both phases ($\approx 300$ kHz), which is less than the distance between any two harmonics for any \mic clock frequency. 

\subsubsection{Candidate Clock Frequency ID}
\label{sec:des-clk-candidate-clk-id}
Given the set of frequency peaks, $\mathbb{F}_{p}$, we predict a list of candidate clock frequencies, $\mathbb{F}_{c}$, and their corresponding set of harmonics, $\mathbb{H}_{c}$. Recall from \S\ref{sec:bg-clock} that clock signals consist of a fundamental frequency (1x), and harmonics (2x, 3x, etc.), as peaks in the frequency domain. Hence, for their robust identification, we require detection of a minimum number (i.e., $\theta_{h}$) of their harmonics (inclusive of the fundamental frequency). By doing so, we prevent prediction of spurious clock signals. 

This is straightforward if we assume that fundamental frequency is always detected as a peak. We can iteratively check the likelihood of each frequency peak, $f_i$, to be a candidate clock frequency. For example, consider the peaks identified in Figure~\ref{fig:des-clk-steps}(d-1), where we compute the likelihood of the first peak (denoted by $\bigstar$), to be a clock frequency. We observe that it has a total of \textit{eight} harmonics (1x, 3x, \dots, 15x), hence, frequency, $f_1$, would be added to the set of candidate clock frequencies, $\mathbb{F}_{c}$ (default value of threshold, $\theta_h = 4$). 

However, the aforementioned approach 
does not work if the fundamental frequency is missing (see Figure~\ref{fig:des-clk-steps}(d-2)). In fact, as we show later in \S\ref{sec:eval-overall}, more than 60\% of the EM traces have a missing fundamental frequencies. Hence, we check the likelihood of each peak to not only be the fundamental frequency, but also a \textit{harmonic} of a potential clock frequency. For example, for the first peak (denoted by $\blacktriangle$) in Figure~\ref{fig:des-clk-steps}(d-2), we check for its likelihood to be a third harmonic. By doing so, we indirectly check for the likelihood of the missing fundamental ($= \frac{f_1}{3}$) to be candidate clock frequency. 
In general, we check for each peak's likelihood to be one of the first `H' harmonics ($H = 10$), thereby handling the case of not just the missing fundamental, but also its several harmonics. Finally, this module outputs the set of candidate clock frequencies, $\mathbb{F}_{c}$, and their corresponding set of detected harmonics, $\mathbb{H}_{c}$. 

\subsubsection{Clock Frequency Pruning}
\label{sec:des-clk-pruning}
We prune the set of candidate clock frequencies, $\mathbb{F}_{c}$, by leveraging their harmonics, $\mathbb{H}_c$, to obtain the final set of clock frequencies, $\mathbb{F}$, and their harmonics, $\mathbb{H}$. 
We identify frequency pairs, $(f_1, f_2)$, both belonging to the candidate set, $\mathbb{F}_{c}$, such that -- (1) the set of harmonics of one is a proper subset of the other (i.e., $\mathbb{H}_c(f_2) \subset \mathbb{H}_c(f_1)$); or (2) the set of harmonics is identical (i.e., $\mathbb{H}_c(f_1) = \mathbb{H}_c(f_2)$). In both these cases, we eliminate one of the two frequencies (i.e., $f_1$ or $f_2$). 
Figure~\ref{fig:des-clk-steps}(e) depicts an example for case (1), where the frequency pair consists of candidate frequencies, $f_1 = 2.048$ MHz, and $f_2 = 6.144$ MHz, where $f_2 = 3.f_1$. Clearly, the harmonics of $f_2$, are a subset of the harmonics of frequency, $f_1$, hence we prune the frequency, $f_2$, which is likely a spurious prediction.
As an example for case (2), we consider frequencies, $f_1=2.048$ MHz, and $f_2 = 1.024$ MHz. All the harmonics of frequency, $f_1$, are also harmonics of the frequency, $f_2$ as, $f_2 = \frac{f_1}{2}$ (e.g., 3x of $f_1$ is 6x of $f_2$). 
We prune the smaller frequency, $f_2$, 
as had it been the underlying clock frequency, we would expect to detect intermediate harmonics (e.g., 3.072 MHz) of frequency, $f_2 = 1.024$ MHz, that are \textit{not} harmonics of frequency, $f_1 = 2.048$ MHz. 
Finally, after pruning the spurious frequencies, 
we output the retained clock frequencies, denoted by $\mathbb{F}$. Optionally, we also output the \textit{number of detected harmonics} 
($\#\mathbb{H}$), 
as required in \S\ref{sec:des-probing_location_id}.

\begin{figure}[!t]
        \centering
        \includegraphics[width=0.9\linewidth]{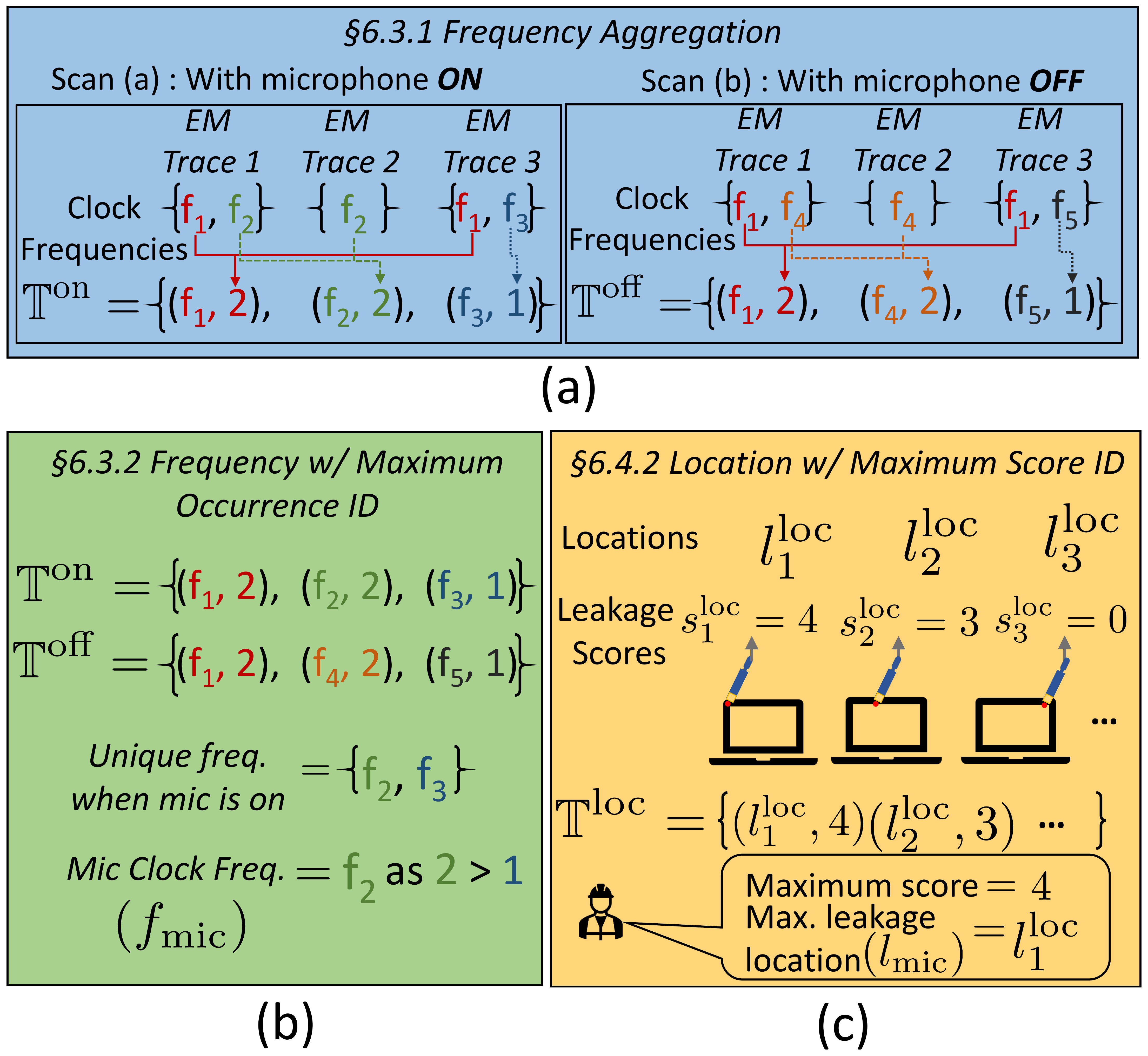}
        \caption{(a) depicts frequency aggregation where we combine clock frequencies obtained from multiple EM traces into a set of tuples ($\mathbb{T}^\textrm{on}/\mathbb{T}^\textrm{off}$), consisting of unique frequencies and their count. (b) depicts how we determine the \mic clock frequency, ~\fclk, from  $\mathbb{T}^\textrm{on}$, and $\mathbb{T}^\textrm{off}$. (c) depicts how we choose the maximum leakage location, \lclk, by identifying the location with the maximum leakage score.}
        \label{fig:des-modules}
\end{figure}
%
\subsection{Challenge 2: \Mic Clock Frequency (\fclk) ID}
\label{sec:des-mic_freq_id}
As part of the \textit{Bootstrapping} phase, we identify the \mic clock frequency, \fclk. The main challenge, however, is that its value is device dependent, particularly on the clock frequencies supported by the device audio hardware, hence is not known a priori. 
To circumvent this problem, we identify \fclk~ by taking as input the EM leakage signals captured from two scans -- \textit{Scan (a)} when the \mic is \textit{on}, and \textit{Scan (b)} when the \mic is \textit{off} (see Figure~\ref{fig:des-fig}(a)). Subsequently, we collect a total of $n_a$ and $n_b$ EM traces, respectively, across different locations around the \mic (e.g., laptop's top bezel). Although the scans are performed over the same region, the number of traces, $n_a$ and $n_b$, can be different. We input each of these traces to the \textit{Clock Frequency Detection} module (\S\ref{sec:des-clk_freq_detection}) to obtain the set of clock frequencies per trace. Note that the number of clock frequencies output can be zero or more, depending on the precise location where the EM trace is captured. 

\subsubsection{Frequency Aggregation}
\label{sec:des-freq-aggr}
We now combine the frequencies obtained from all the traces of a particular scan, and output a set of tuples, containing the unique frequencies present and their number of occurrences (i.e., count). In particular, for \textit{Scan (a)},  
we obtain the set of tuples, $\mathbb{T}^{\textrm{on}}$, consisting of -- $\{(f^\textrm{on}_1,c^\textrm{on}_1), (f^\textrm{on}_2,c^\textrm{on}_2), \dots\}$, where frequency, $f^\textrm{on}_i$, indicates a distinct frequency, and the count, $c^\textrm{on}_i$, indicates the total number of occurrences of the frequency, $f^\textrm{on}_i$. 
Likewise, for \textit{Scan (b)}, 
we obtain the set of tuples, $\mathbb{T}^{\textrm{off}}$. Figure~\ref{fig:des-modules}(a) depicts this with a toy example with three EM traces. 
Furthermore, among the tuples of a single scan, we merge frequencies that are within an error margin ($\theta_{margin} \approx 10$ kHz) of each other into a single frequency, by summing up their individual count values. 

\subsubsection{\Mic Frequency with Maximum Occurrence Identification}
\label{sec:des-mic-freq-max}
This module takes as input, the set of tuples, $\mathbb{T}^\textrm{on}$, and $\mathbb{T}^\textrm{on}$, obtained from \textit{Frequency Aggregation} module, in order to output the \mic clock frequency, \fclk. 
Figure~\ref{fig:des-modules}(b) depicts how we first identify the frequencies that \textit{uniquely} occur in set, $\mathbb{T}^\textrm{on}$ (and hence absent in the set, $\mathbb{T}^{\textrm{off}}$). Subsequently, we choose \fclk~ as the one with the \textit{maximum} count value among all the unique frequencies (in cases with more than one unique frequency). 

We also identify the average leakage amplitude, corresponding to \fclk, by computing the average amplitude of the \mic clock signals (i.e., clock frequency and harmonics), in traces where \fclk~ is detected. The average leakage amplitude differs across devices, hence is leveraged as a threshold ($\theta_a$) for successful detection of \fclk~ in the \textit{Clock Frequency Detection} module of all the subsequent stages (see \S\ref{sec:des-clk-peak-detection}). 

At the end of this step, if we fail to identify any unique clock frequency across different scanning locations (e.g., top bezel, bottom bezel, and so on), we conclude that \name's technique does not hold for such a device.

\subsection{Challenge 3: Max Leakage Location (\lclk) ID}
\label{sec:des-probing_location_id}
This module which is part of the \textit{Bootstrapping} phase, takes as input the EM signals along with their \textit{location information}, in order to identify the maximum leakage location, \lclk, corresponding to the \mic clock frequency, \fclk, and its harmonics. 

The main challenge in identifying the EM leakage location is its dependence on the location of underlying leakage sources (e.g., connectors and cables), which in-turn depends on the device's hardware layout. Additionally exacerbating the problem, the leakage region can be highly localized, i.e., to an area as small as a few ${cm}^2$. 
 
In order to identify \lclk, we perform a third scan (\textit{Scan (c)}), with the \mic \textit{on}, along the same scanning region as in \textit{Scans (a) and (b)} (Figure~\ref{fig:des-fig}(a)).  We input each EM trace captured at each location, $l_i^\textrm{loc}$ (\textit{loc} to denote location ID step), to the \textit{Clock Frequency Detection} module (\S\ref{sec:des-clk_freq_detection}), and obtain the set of clock frequencies, $\mathbb{F}^\textrm{loc}_i$, as well as the \textit{number of harmonics} detected per clock frequency, $\#\mathbb{H}^\textrm{loc}_i$.

\subsubsection{Leakage Score Computation}
\label{sec:des-leakage-score}
This module takes as input -- detected clock frequencies, $\mathbb{F}^\textrm{loc}_i$, their corresponding number of harmonics, $\#\mathbb{H}^\textrm{loc}_i$, as well as the identified \mic clock frequency, ~\fclk, to output a leakage score, 
$s_i^\textrm{loc}$. 
We compute the leakage score as the \textit{number of detected harmonics} of \fclk~ obtained from the list, $\#\mathbb{H}^\textrm{loc}_i$. Hence, a location with higher number of detected harmonics for frequency, ~\fclk, has a higher leakage score. However, if \fclk~ is not detected, or if it is detected \textit{in addition} to other spurious frequencies, we output a leakage score of zero, to indicate the unsuitability of the location for reliable detection of \fclk.   

\subsubsection{Location with Maximum Score Identification}
\label{sec:des-loc-max-score}
This module takes as input the set of location and leakage score tuples, i.e., $\{(l_1^\textrm{loc},s_1^\textrm{loc}), (l_2^\textrm{loc},s_2^\textrm{loc}), \dots \}$, to output \lclk, with the \textit{maximum} leakage score, as the best location for probe placement. In the current implementation of \name, this module is performed manually, where-in the person performing the \textit{Bootstrapping} process decides the best location, by probing several locations, and identifying the location with maximum score as provided by our system (see Figure~\ref{fig:des-modules}(c)). However, we highlight that this is a one-time effort, as the \textit{Bootstrapping} phase is performed only once per device. We refer interested readers to -- \url{https://bit.ly/3w2QTDA} for a video demo on how we user perform this scan. 

In general, there could be more than one location with maximum leakage, in which case we choose any one of them as \lclk. On the flip side, if we encounter a device with no suitable locations (e.g., with a score of zero everywhere), this implies that we identified a spurious frequency as \fclk~ in the previous step (\S\ref{sec:des-mic_freq_id}), and hence conclude that \name's approach is inapplicable to such a device.  

\section{Evaluation}
\label{sec:evaluation}
We evaluate \name comprehensively on several devices and for various differing conditions, to demonstrate its feasibility.  
%
\begin{figure}[!t]
     \centering
     \begin{subfigure}[b]{0.5\linewidth}
         \centering
         \includegraphics[width=\textwidth]{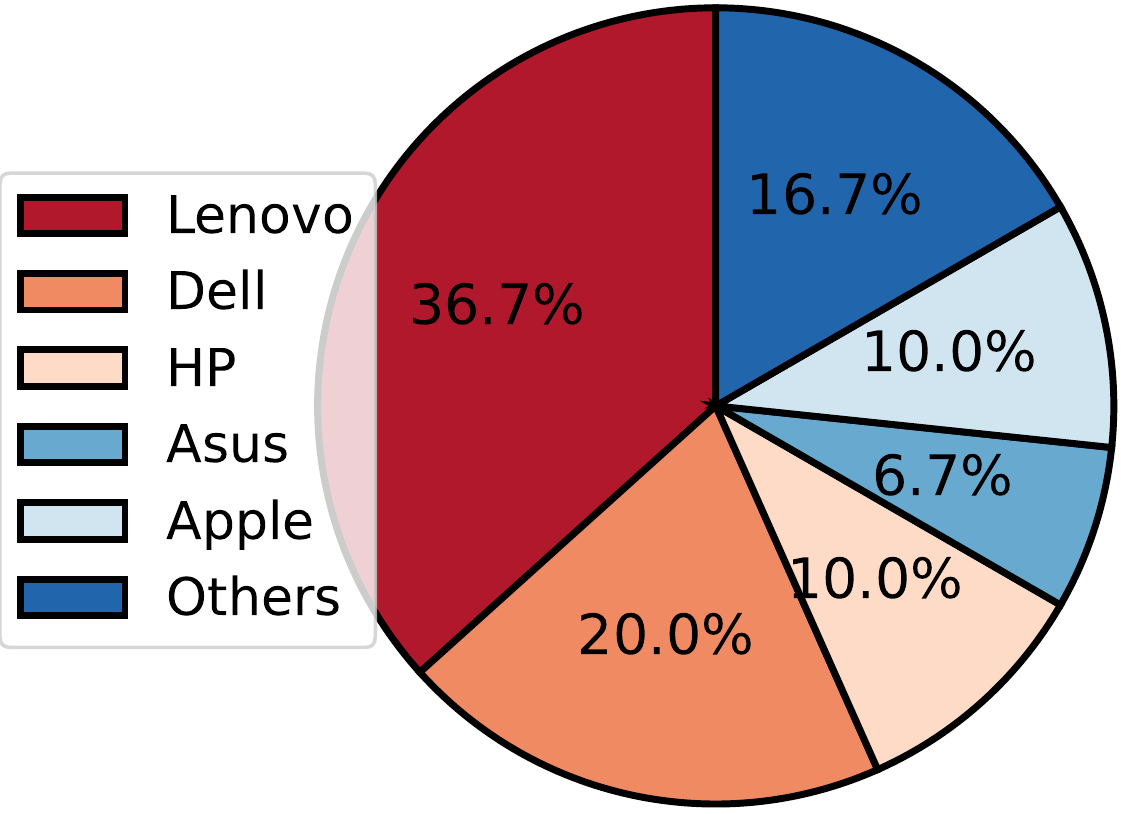}
         \caption{}
         \label{fig:eval-laptop-models}
     \end{subfigure}
     \hfill
     \begin{subfigure}[b]{0.4\linewidth}
         \centering
         \includegraphics[width=\textwidth]{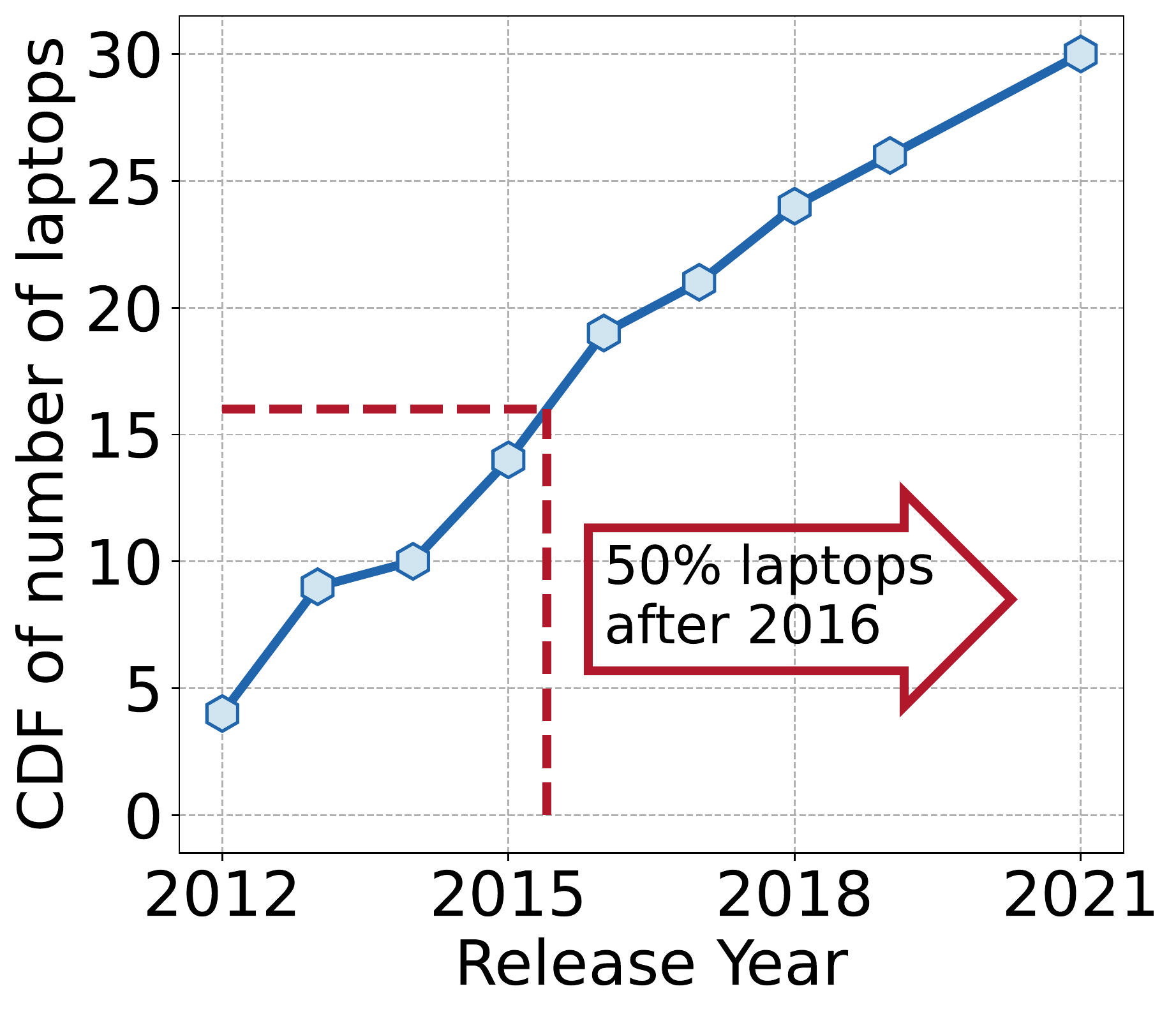}
         \caption{}
         \label{fig:eval-laptop-years}
     \end{subfigure}
     \caption{Figure (a) depicts the brands of the 30 laptops we evaluate, and (b) depicts the release years of the laptops.}
     \label{fig:eval-laptop-data}
\end{figure}
%
\begin{figure}[!t]
        \centering
        \includegraphics[width=0.9\linewidth]{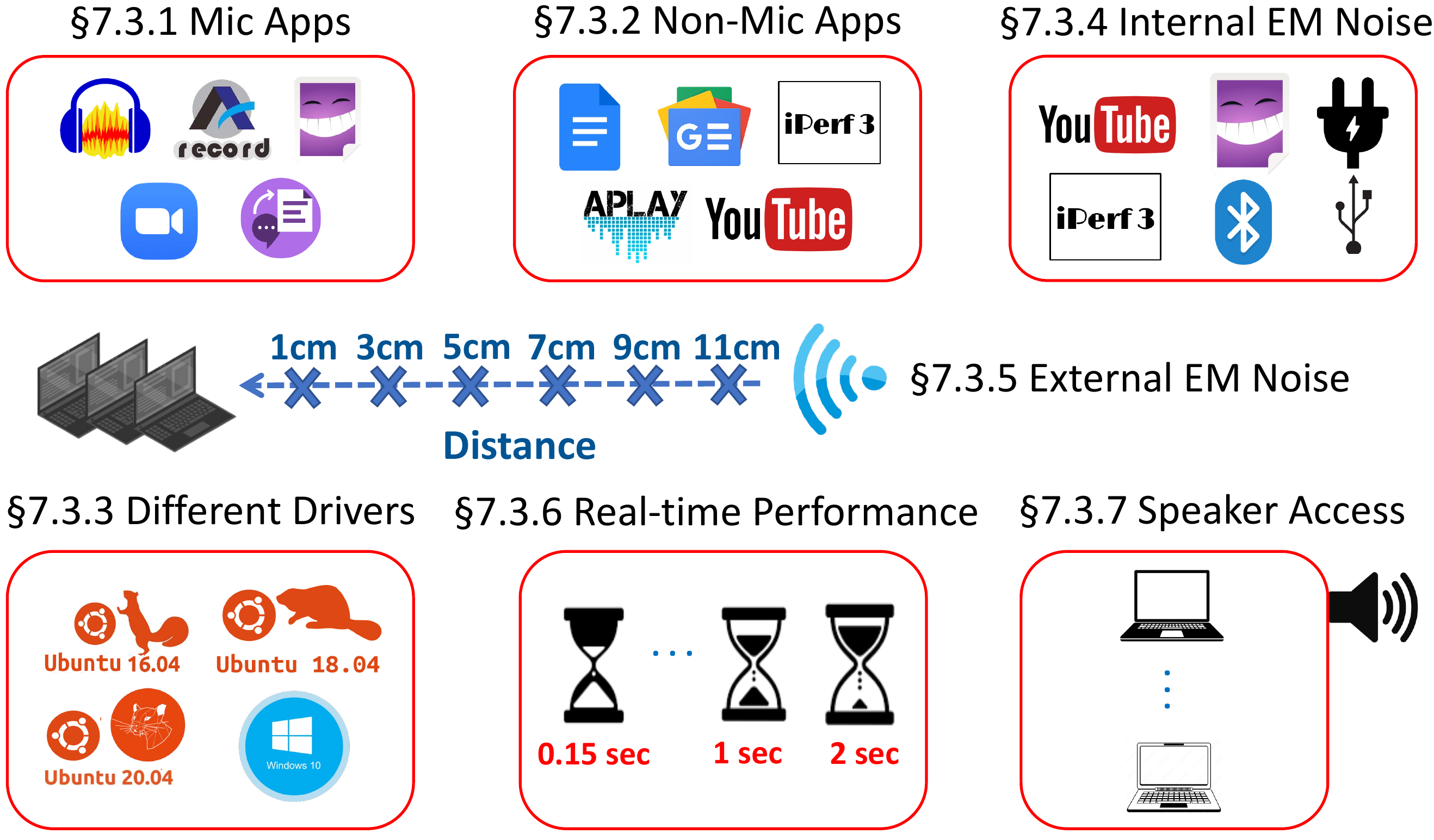}
        \caption{Figure depicts the setup of \name's experiment conditions, specifically when varying different conditions for comprehensive evaluation.
        } 
        \label{fig:eval-diff-exp-conditions}
\end{figure}
\subsection{Experimental Setup}
\label{sec:eval-setup}
\textbf{Apparatus.} 
We utilize the same setup described in \S\ref{sec:feasibility-setup}, 
consisting of the device to be tested (e.g., laptop), a near-field probe, RF amplifier, software defined radio (SDR) and an RPi 4B, for our experiments (see Figure~\ref{fig:prototype}). We test each laptop using both the near-field probes, namely the E-field and H-field probes. We also custom-design an amplifier with gain of 27 dB for its low power consumption~\cite{rf-lna-minicircuits}. 
We leverage RSP-1A SDR (US$\$140$) that captures signals covering a large portion of radio spectrum, from 1 kHz to 2 GHz, with a maximum bandwidth of 10 MHz~\cite{sdrplay}. During the detection process, we sweep across four (overlapping) frequency bands to obtain a total bandwidth of 30 MHz (from $0.85 - 30.85$ MHz) in order to detect mic clock frequencies and their harmonics. 

\noindent\textbf{Data Collection.} 
We evaluate \name on a total of 30 laptops of popular brands including Lenovo, Dell, HP and Apple, all released in the last ten years (see Figures~\ref{fig:eval-laptop-data}(a) and ~\ref{fig:eval-laptop-data}(b)). 
For consistency of experiments, we run Ubuntu 20.04 LTS with kernel version 5.11.0-27 on each of the laptops (except Macbooks that run Mac OS X). We record audio at 32-bit 48 kHz using the command-line tool, \texttt{arecord}, unless mentioned otherwise~\cite{arecord}. 
Furthermore, we ensure that the laptop is plugged into power source and that its screen is active throughout. 
Furthermore, as depicted in Figure~\ref{fig:eval-diff-exp-conditions}, we evaluate \name's performance across different \mic ($\S$\ref{sec:eval-mic-app}) and non-\mic applications ($\S$\ref{sec:eval-non-mic-app}), different audio driver implementations ($\S$\ref{sec:eval-audio-driver}), its robustness to internal and external EM noise ($\S$\ref{sec:eval-internal-em-noise}, $\S$\ref{sec:eval-em-noise}), its real-time performance ($\S$\ref{sec:eval-real-time}), the influence due to speaker-access ($\S$\ref{sec:eval-speaker}), as well as the effect of varying sound levels (Appendix~\ref{app:eval-sound-levels}). 
For the evaluation, we perform \name's detection in an offline manner, i.e., we identify the clock signals \textit{after} capturing all traces (except in \S\ref{sec:eval-real-time}). Furthermore, we determine the number of harmonics to be identified (i.e., parameter, $\theta_h$) to be \textit{three}, based on the results from Appendix~\ref{app:module-eval}. 

\noindent\textbf{Performance Metrics.} 
We define the following three metrics to evaluate \name's overall results. 
\textit{Device Hit Rate} refers to the fraction of total devices tested in which \name identifies the \mic clock frequency, \fclk. Furthermore, we leverage \textit{True Positive Rate (TPR)} and \textit{True Negative Rate (TNR)} to evaluate the performance of ~\name in predicting \mic status (\textit{on} vs \textit{off}) in devices. 
We consider an EM trace to be a \textit{positive} example, if \name detects \fclk~ as the \textit{only} clock frequency from the EM trace (and \textit{negative} example otherwise). Hence, we define TPR as the fraction of all traces that are identified to be \textit{positive} examples, when the mic is \textit{on}, and TNR as the fraction of all traces that are identified as \textit{negative} examples, when the mic is \textit{off}.

\begin{table*}[!t]
\begin{tabular}{|l|l|l|l||l|l|l|l||l|l|l|l|}
\hline
Device Model & \multicolumn{1}{l|}{\begin{tabular}[c]{@{}l@{}}\fclk \\ (kHz)\end{tabular}} & \multicolumn{1}{c|}{\begin{tabular}[c]{@{}c@{}}Unique\\Clk?\end{tabular}} & A/D & Device Model & \multicolumn{1}{l|}{\begin{tabular}[c]{@{}l@{}}\fclk \\ (kHz)\end{tabular}} & \multicolumn{1}{c|}{\begin{tabular}[c]{@{}c@{}}Unique\\Clk?\end{tabular}} & A/D & Device Model & \multicolumn{1}{l|}{\begin{tabular}[c]{@{}l@{}}\fclk \\ (kHz)\end{tabular}} & \multicolumn{1}{c|}{\begin{tabular}[c]{@{}c@{}}Unique\\Clk?\end{tabular}} & A/D \\ \hline
ASUS Strix & 2048 & \cmark & D & HP Probook 440 G1 & 2352 & \cmark & D & Lenovo X230 & 2048 & \cmark & D \\ \hline
Asus X450v & 2048 & \cmark & D & HP Zbook Studio G5 & 3072 & \cmark & D & Lenovo X250 & 2048 & \cmark & U \\ \hline
Dell Inspiron 13 & 3072 & \cmark & D & Lenovo P14s gen 1 & 2400 & \cmark & U & Lenovo X260 & 2048 & \cmark & D \\ \hline
Dell Inspiron 5459 & 2048 & \cmark & D & Lenovo T430U & 2048 & \cmark & D & Razer RZ09-0102 & 2048 & \cmark & D \\ \hline
Dell Inspiron 7572 & 2048 & \cmark & D & Lenovo T460s & 2048 & \cmark & U & Samsung Chronos & 2048 & \cmark & U \\ \hline
Dell Latitude E5570 & 2048 & \cmark & U & Lenovo T470S & 3072 & \cmark & U & Terrans Force T5 & 2048 & \cmark & U \\ \hline
Dell Latitude E7450 & 2048 & \cmark & U & Lenovo T590 & 2048 & \cmark & U & Toshiba Portege & 6144 & \cmark & U \\ \hline
Dell XPS L321x & 2048 & \cmark & D & Lenovo X1 Carbon G7 & 2400 & \cmark & U & Mac Pro 2014 15” & \multicolumn{1}{l|}{--} & {\color{red}\xmark} & U \\ \hline
Fujitsu Lifebook & 2048 & \cmark & D & Lenovo X1 Extreme G3 & 2400 & \cmark & U & Mac Pro 2017 13” & 2823 & {\color{brown}$\otimes$} & U \\ \hline
HP Envy 13 & 3072 & \cmark & D & Lenovo X13 Gen 2 & 2400 & \cmark & U & Mac Pro 2019 16” & \multicolumn{1}{l|}{--} & {\color{red}\xmark} & U \\ \hline
\end{tabular}
\caption{Evaluation of \name on a total of 30 laptops. \name can successfully detect \mic clock frequency, \fclk, on 27 out of the 30 laptops, i.e., 90\% of all tested laptops. \cmark~depicts successful, {\color{red}\xmark}~depicts unsuccessful detection, and {\color{brown}$\otimes$} depicts confounding cases. The A/D column indicates whether the microphone(s) in the laptops are digital (D), analog (A) or unknown (U). We present more detailed results in Appendix~\ref{app:laptop-results}.}
\label{tbl:eval-laptop}
\end{table*}
%
\subsection{\name Performance}
\label{sec:eval-overall}
We present \name's overall performance by first presenting the results of bootstrapping followed by their performance in determining \mic's \textit{on/off} status. 

\begin{figure}[!t]
     \centering
     \includegraphics[width=0.75\linewidth]{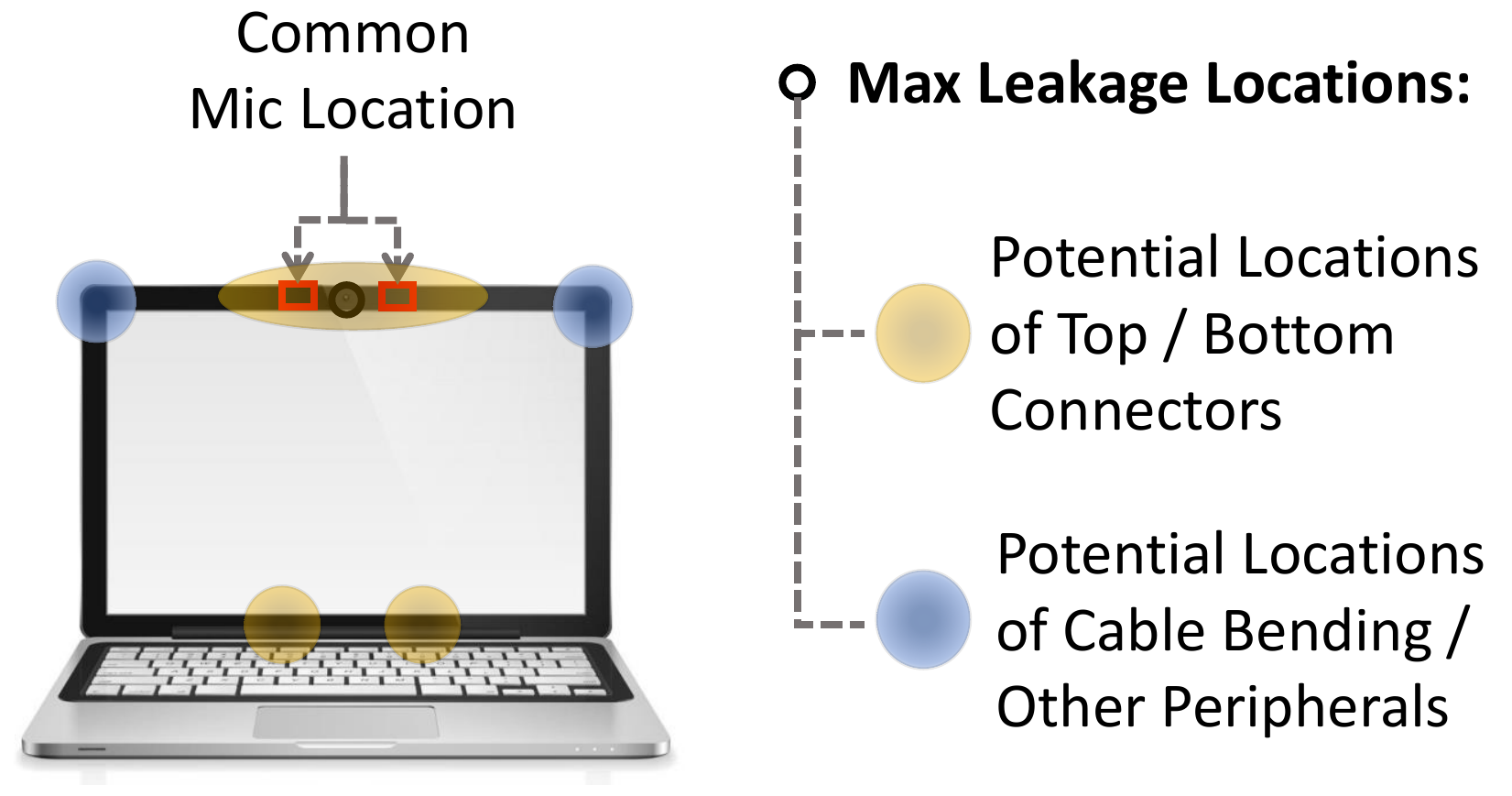}
     \caption{Figure depicts regions containing maximum leakage location across laptops having their \mics on either sides of the webcam on the laptop's top bezel. We believe that leakage at locations within regions highlighted in yellow are due to connectors, while those within the blue regions are due to cable bending at corners or presence of other peripherals. Please refer to the link -- \url{https://bit.ly/3kAqpnH} for images of leakage locations for each of the 27 laptops.}
     \label{fig:eval-mic-leakage}
\end{figure}
%
\subsubsection{Bootstrapping Summary}
\label{sec:eval-bootstrapping-summary}
In Table~\ref{tbl:eval-laptop}, we present 30 laptops we test, along with their detected \fclk, if any, from the bootstrapping phase. We achieve a \textit{device hit rate} of 90\%, as we successfully identify \fclk~ for 27 laptops, with \fclk~ ranging between $2.048 - 6.144$ MHz.
Figure~\ref{fig:eval-mic-leakage} depicts prominent leakage locations, \lclk, observed on laptops that 
have their \mics located on either sides of the webcam on the top bezel.\footnote{We also present detailed evaluation results in Appendix~\ref{app:laptop-results} including information such as the leakage amplitude, performance due to each near-field probe and the harmonics detected per laptop. }
The leakage locations within regions annotated in yellow potentially correspond to locations of connectors (either top/bottom), while those annotated in blue represent potential locations of cable bends or presence of other peripherals (see \S\ref{sec:bg-factors}). 

Although our approach works well on 90\% of the tested laptops, including \textit{all} tested models from popular vendors such as Lenovo, Dell, HP and Asus, \name fails to detect the \mic clock signals in three laptops, all of which are Apple Macbooks. On each of the tested Macbooks, the \mics are located either on the left or right side of the keyboard (along the speaker vent), and are connected to the motherboard via short flex cables. We believe that the aluminium enclosure of Macbooks, along with the usage of short flex cables, result in significantly attenuating the leakage signal~\cite{mac-14, mac-17, mac-19, mac-al}. Of the three laptops, we encounter a confounding case in Macbook Pro 2017 (13"), where a \mic clock frequency (\fclk~$=2.823$ MHz) with a low leakage amplitude is detected, although its detection fails to be consistent across different audio recording applications (i.e., clock frequency is absent for some audio recording apps but present for others). 
\revision{We tested ten additional Macbooks using a different setup consisting of a high gain amplifier and spectrum analyzer. However, \name is still unable to detect the clock signals consistently. The results are shown in the Appendix \ref{app:sec-macbook}.}

\begin{figure}[!t]
     \centering
     \begin{subfigure}[b]{0.4\linewidth}
         \centering
         \includegraphics[width=\textwidth]{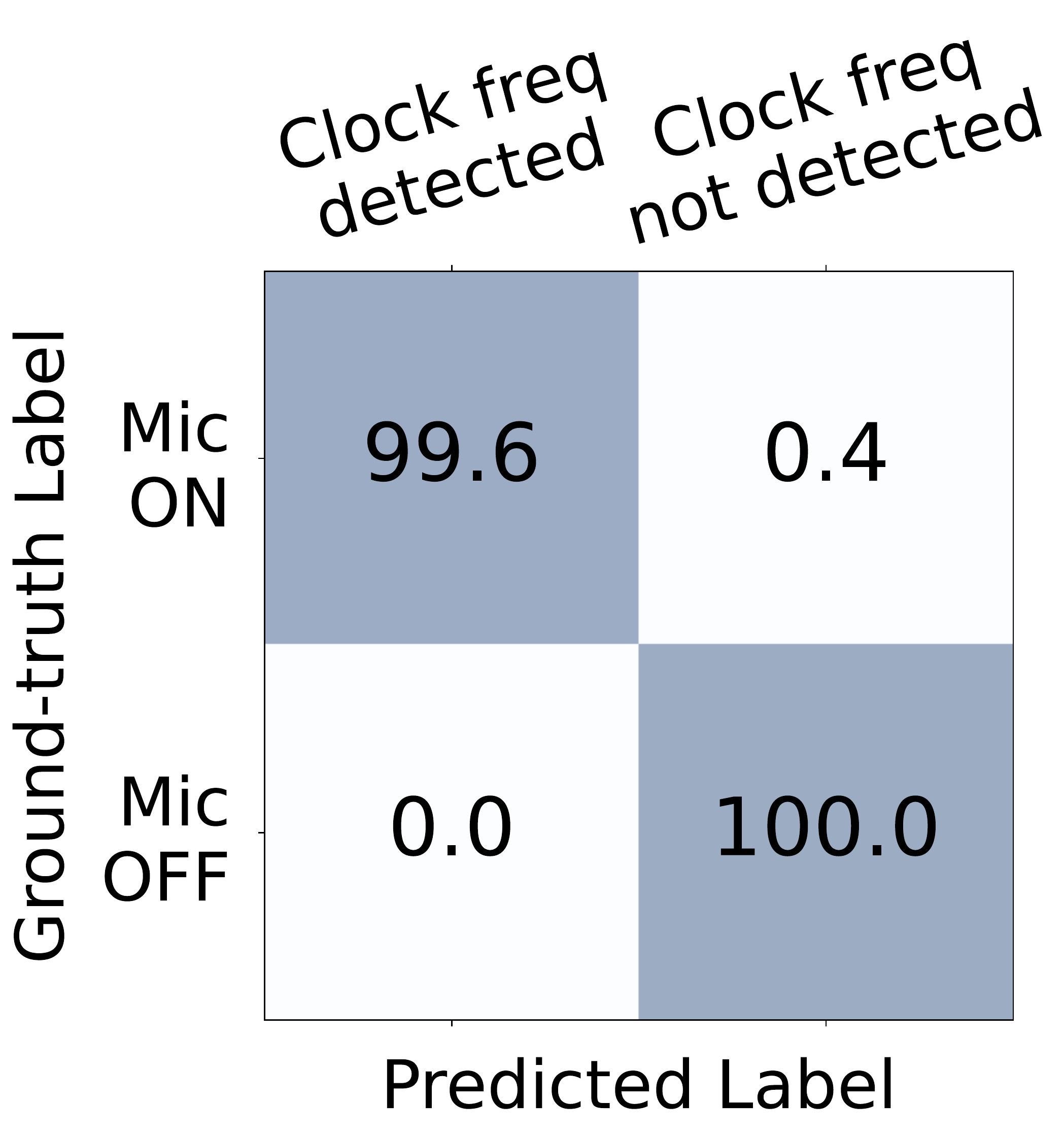}
         \caption{}
         \label{fig:eval-conf-matrix}
     \end{subfigure}
     \hfill
     \begin{subfigure}[b]{0.5\linewidth}
         \centering
         \includegraphics[width=\textwidth]{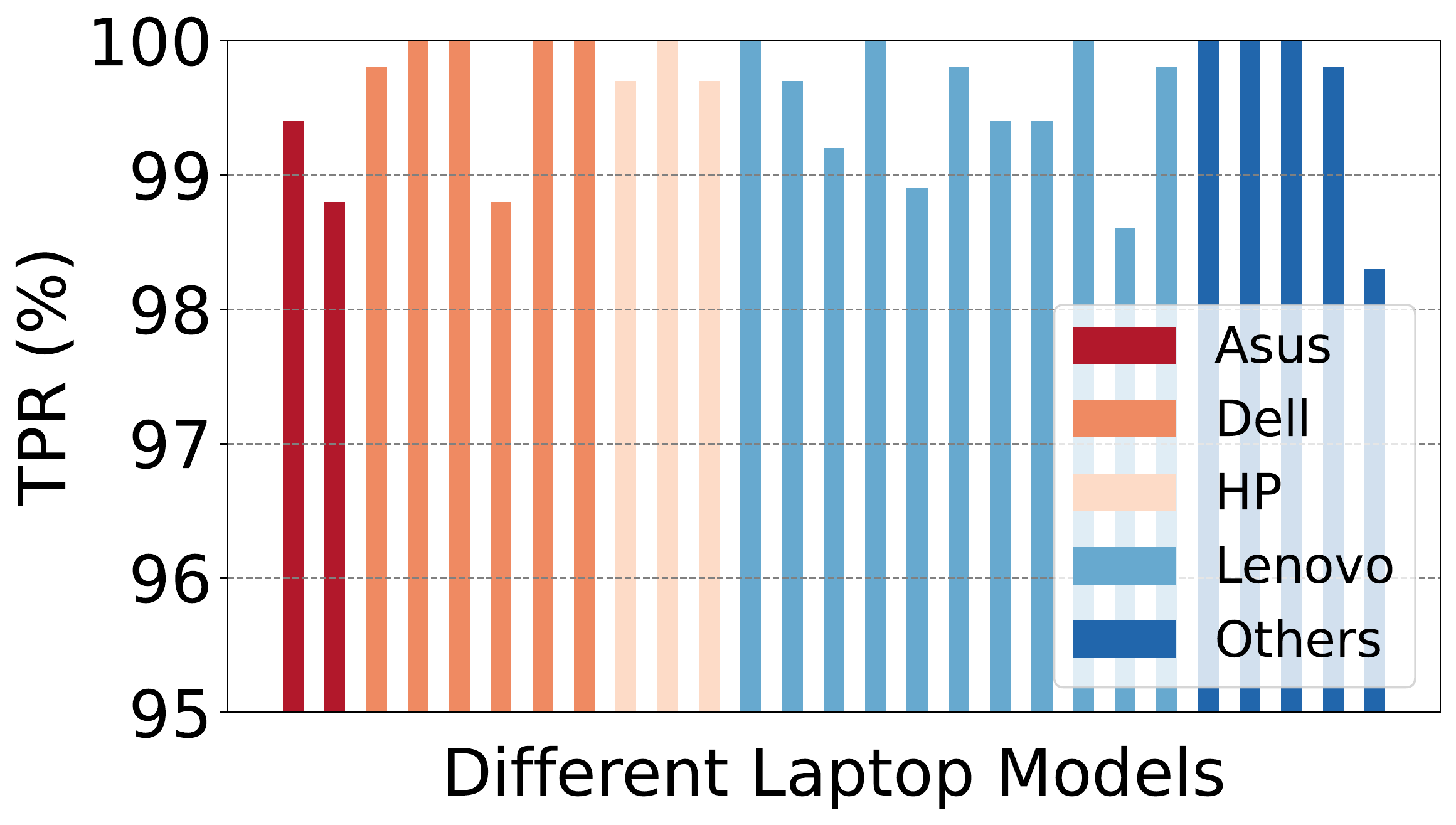}
         \caption{}
         \label{fig:eval-laptop-tpr}
     \end{subfigure}
     \caption{Figure (a) depicts the confusion matrix representing \name's overall detection efficacy, and (b) depicts the individual TPR,  for the 27 successful laptops.}
     \label{fig:eval-confmat-tpr}
\end{figure}
%
\begin{figure}[!t]
         \centering
         \includegraphics[width=0.75\linewidth]{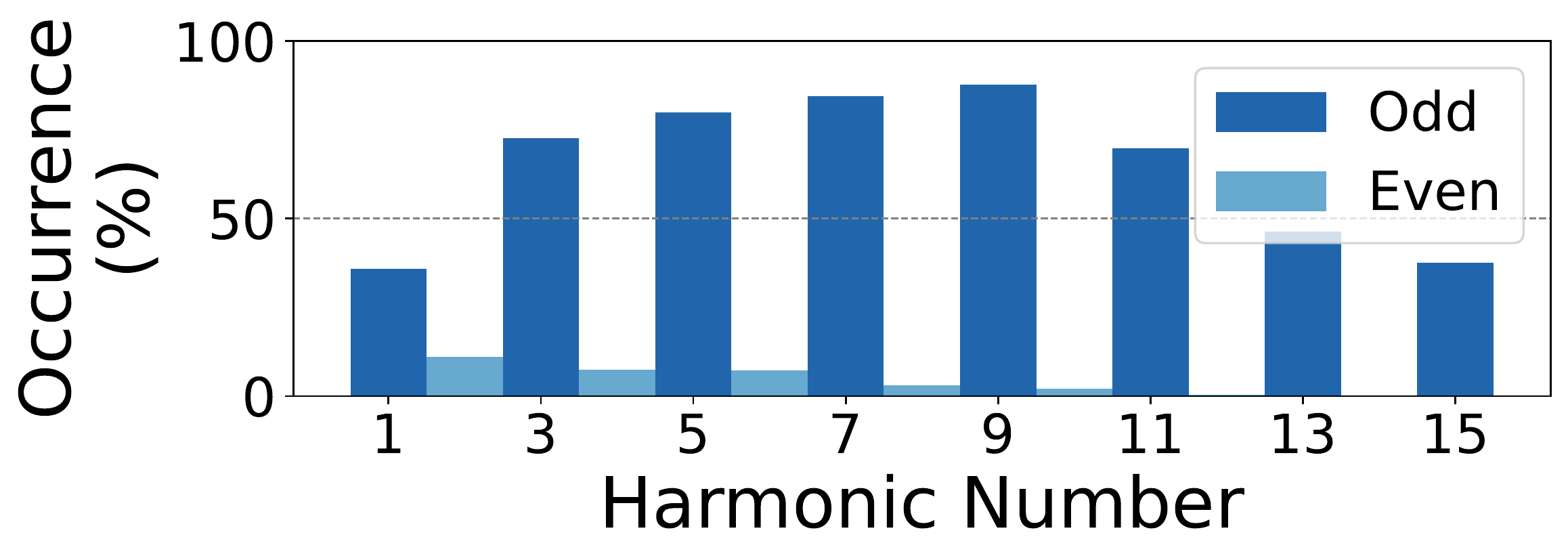}
         \caption{Figure depicts a histogram of the total number of occurrences of different harmonics (as a percentage) among the successfully detected trials across all 27 laptops.}
         \label{fig:eval-hmonics-histogram}
\end{figure}
%
\subsubsection{\Mic Activity Detection Efficacy}
\label{sec:eval-laptop-mic-activity}
In order to verify how \textit{reliable} the clock signals are in identifying \mic activity, we collect EM traces for three minutes at an average rate of 3.6 traces per second (i.e., a total of 650 traces), while -- (1) running an audio recording application (i.e., \texttt{arecord}); and (2) \textit{not} running any audio recording 
application.
We obtain a high true positive rate of 99.6\% (i.e., $\frac{17480}{17550}$), and a 100\% true negative rate (i.e., $\frac{17550}{17550}$) across the 27 successful laptops (see Figure~\ref{fig:eval-confmat-tpr}(a)). Furthermore, the minimum true positive rate across all laptops is above 98\% (see Figure~\ref{fig:eval-confmat-tpr}(b)), depicting the \textit{reliability} of \name as a mic activity indicator.    

Recall from \S\ref{sec:des-clk_freq_detection} that we detect clock frequencies by identifying their harmonics. Hence, in Figure~\ref{fig:eval-hmonics-histogram}, we plot the prominence of different harmonics in the detected clock signals, as a fraction of all traces with successful clock frequency detection (i.e., across $17480$ traces). We observe that the odd harmonics are more prominent compared to even harmonics (as expected from Figure~\ref{fig:bg-clock}). Furthermore, we observe that less than $40\%$ of all successful cases contain the fundamental frequency (i.e., the first harmonic), hence validating our design that accounts for missing fundamentals (see \S\ref{sec:des-clk-candidate-clk-id}).

\subsection{Differing Experimental Conditions}
\label{sec:eval-differing-cond}
We evaluate \name's performance over several factors. For this purpose, we choose three representative laptops, 
Lenovo Thinkpad T430U (\lold), Lenovo Thinkpad T470s (\lmid), and Lenovo X1 Extreme Gen 3 (\lnew), released in 2012, 2017 and 2021, respectively. 
We report our results by capturing 650 EM traces over three minutes per device for each differing condition.  

\begin{figure}[!t]
    \centering
    \includegraphics[width=0.8\linewidth]{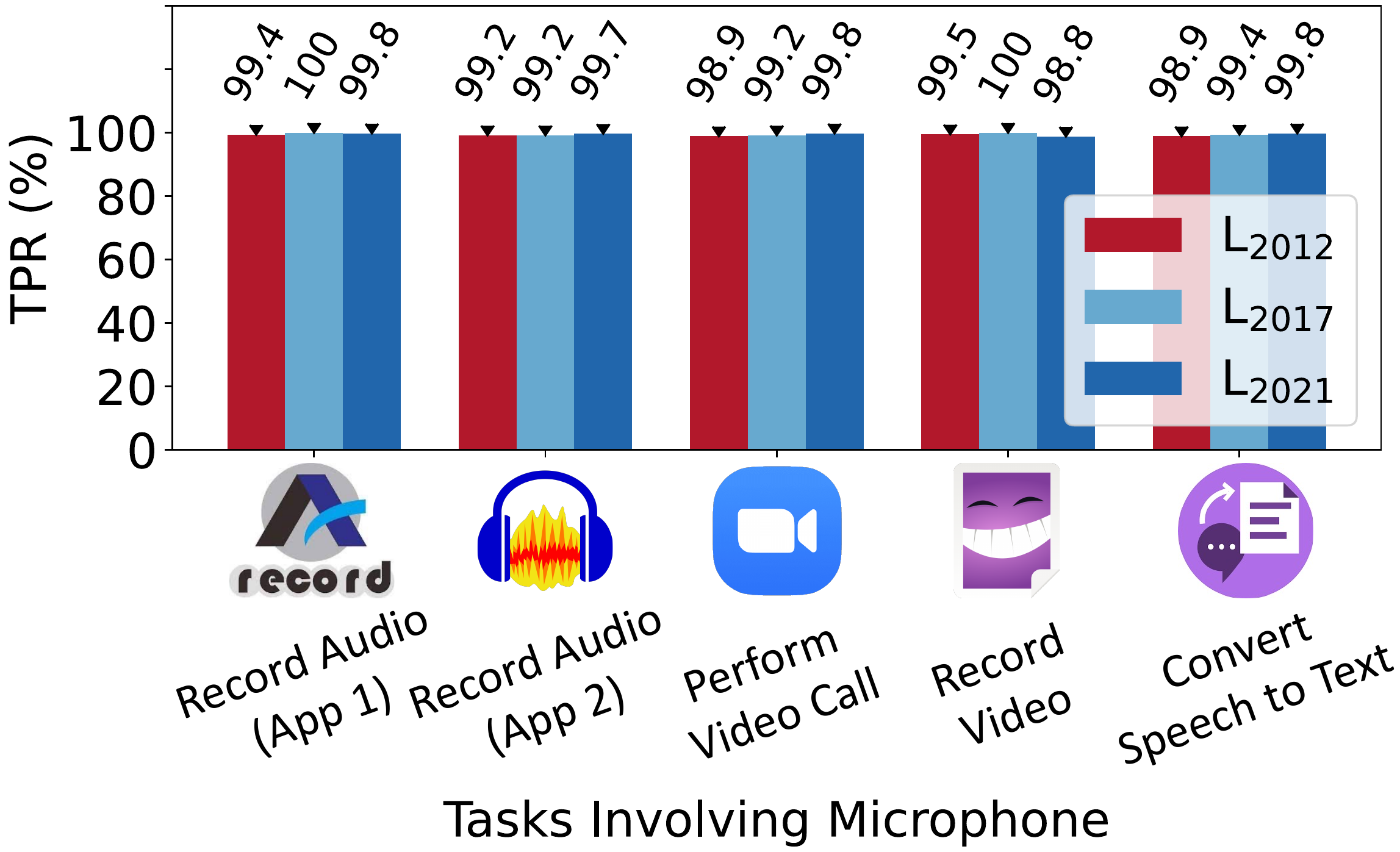}
    \caption{Figure depicts the \textit{consistency} of \name's detection of \mic's \textit{on} status in the presence of several \mic-based applications.} 
    \label{fig:eval-audio-applications}
\end{figure}
%
\subsubsection{Running \Mic-based Applications}
\label{sec:eval-mic-app}
To evaluate \name's performance in detecting \mic \textit{on} status while capturing audio, we report the \textit{true positive rate (TPR)} obtained on running five applications namely -- Audacity and \texttt{arecord} (for recording audio), Zoom (for performing video calls), Cheese (for recording video), and browser-based IBM Watson Speech to Text Service (for transcribing audio). 
\name obtains high TPR for all applications over the three laptops, with a minimum TPR of 98.8\% (i.e., $\frac{642}{650}$) obtained for recording video on laptop,~\lnew~(see Figure~\ref{fig:eval-audio-applications}). These results represent the \textit{consistency} of \name in identifying \mic \textit{on} status. 

\begin{figure}[!t]
    \centering
    \includegraphics[width=0.8\linewidth]{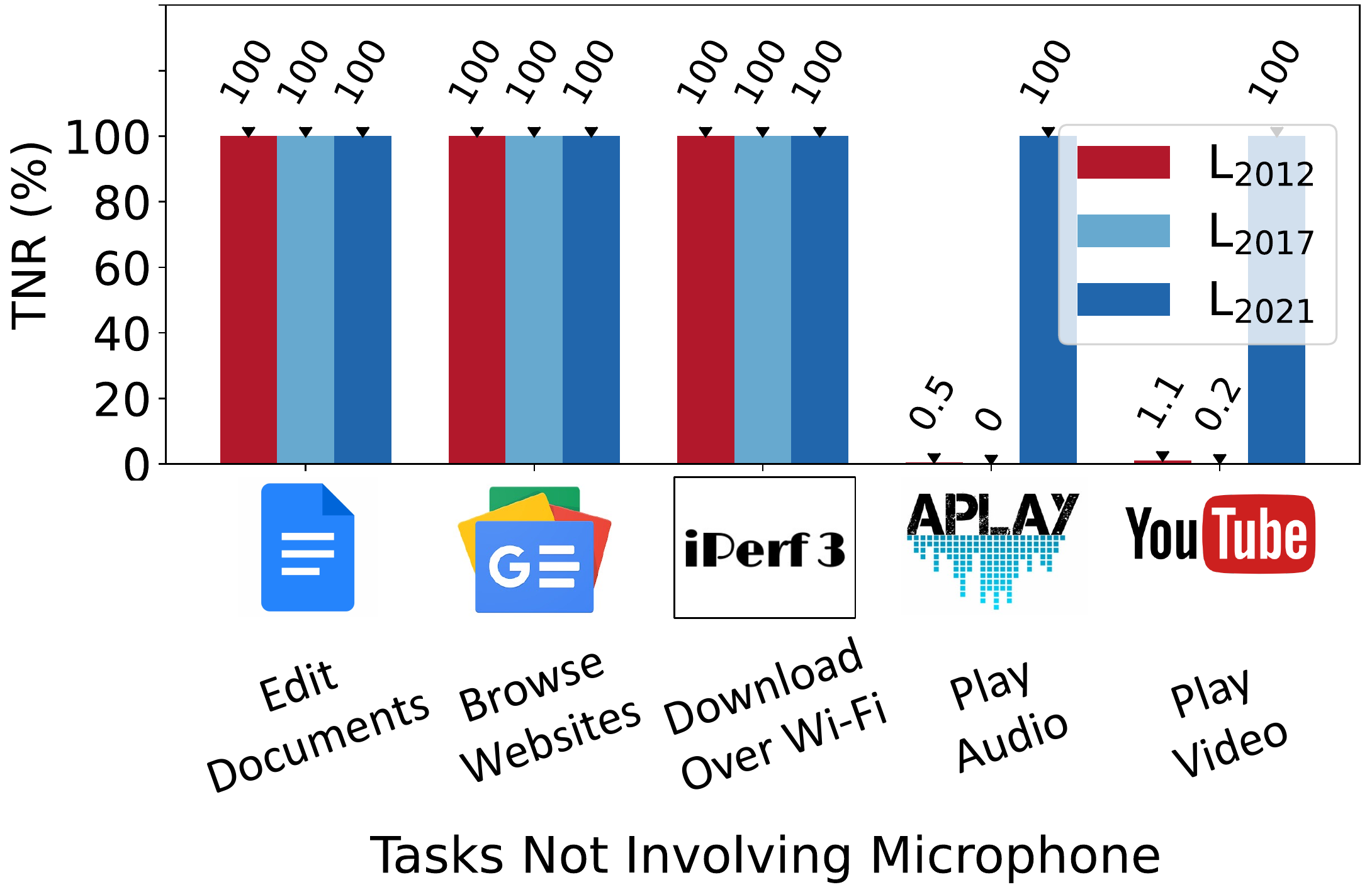}
    \caption{Figure depicts the TNR achieved by \name on running non-\mic based applications.} 
    \label{fig:eval-non-mic}
\end{figure}
%
\subsubsection{Running Non-\mic based Applications}
\label{sec:eval-non-mic-app}
To evaluate false triggers during \mic \textit{off} state, we evaluate \name by performing everyday tasks (that do not involve the \mic) such as taking notes, browsing news, downloading data at high speed (100 Mbps) over Wi-Fi, playing audio and playing video, using five representative applications/tools, namely, Google Docs, Google News, iPerf3, ~\texttt{aplay}, and YouTube, respectively.  
From Figure~\ref{fig:eval-non-mic}(a), we observe that, for the first three tasks, \name obtains a TNR of 100\% across all laptops. However, for the last two applications involving access to speaker, although the newest laptop,~\lnew, continues to achieve 100\% TNR, the older laptops, i.e., ~\lold~and ~\lmid, obtain a TNR of less than 2\%. We defer the explanation of this result to \S\ref{sec:eval-speaker}.

\begin{figure}[!t]
    \centering
    \includegraphics[width=0.9\linewidth]{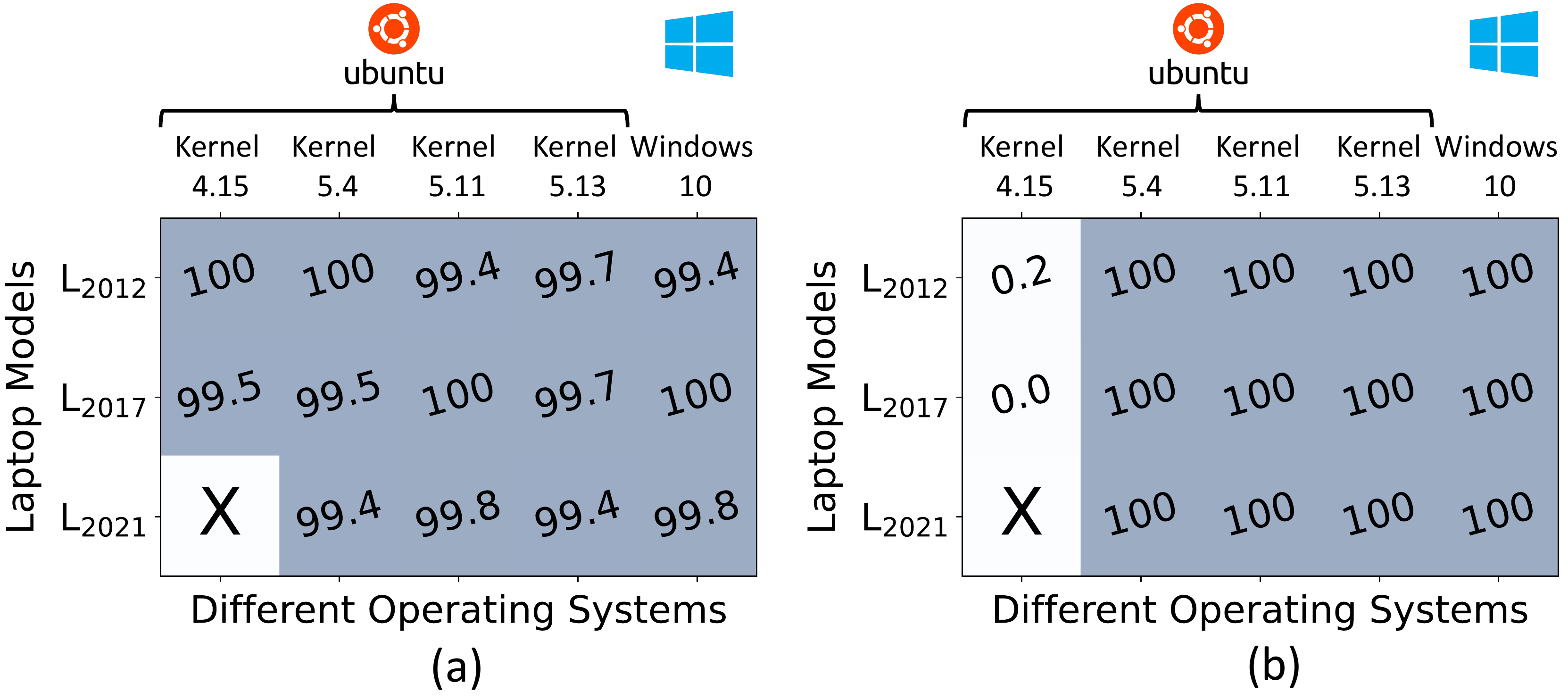}
    \caption{Figure depicts the (a) TPR and (b) TNR obtained for three laptops running five operating systems with different audio driver implementations.} 
    \label{fig:eval-audio-driver}
\end{figure}

\subsubsection{Effect of Different Audio Driver Implementations}
\label{sec:eval-audio-driver}
We evaluate the effect of different audio driver implementations on \name's performance. 
We consider drivers that are part of the OS -- Ubuntu 16.04 (kernel v4.15), Ubuntu 18.04 (kernel v5.4), Ubuntu 20.04 (kernel v5.11 and v5.13), as well as the driver on Windows 10. We evaluate all laptop-OS combinations, with the exception of laptop,~\lnew, with Ubuntu 16.04, due to lack of compatibility (depicted by a $\times$ in the figure).  
As depicted in Figure~\ref{fig:eval-audio-driver}, for all OSes except Ubuntu 16.04, the TPR and TNR are consistently above 99\% across all laptops. However, in the case of Ubuntu 16.04, for laptops, ~\lold, and ~\lmid, although the TPR is above 99\%, the TNR is close to 0\%. On further analysis, we infer that in this driver implementation, the clock signal is \textit{always} provided to the \mic, \textit{irrespective} of whether the \mic is \textit{on} or \textit{off}, resulting in a low TNR. We believe that the future Linux driver implementations' retract clock signals in order to enhance security (i.e., prevent accidental audio capture), while conserving power in laptops. 
Hence, newer driver versions will likely follow suit, thereby improving \name's accuracy. 


\begin{figure}[!t]
    \centering
    \includegraphics[width=0.8\linewidth]{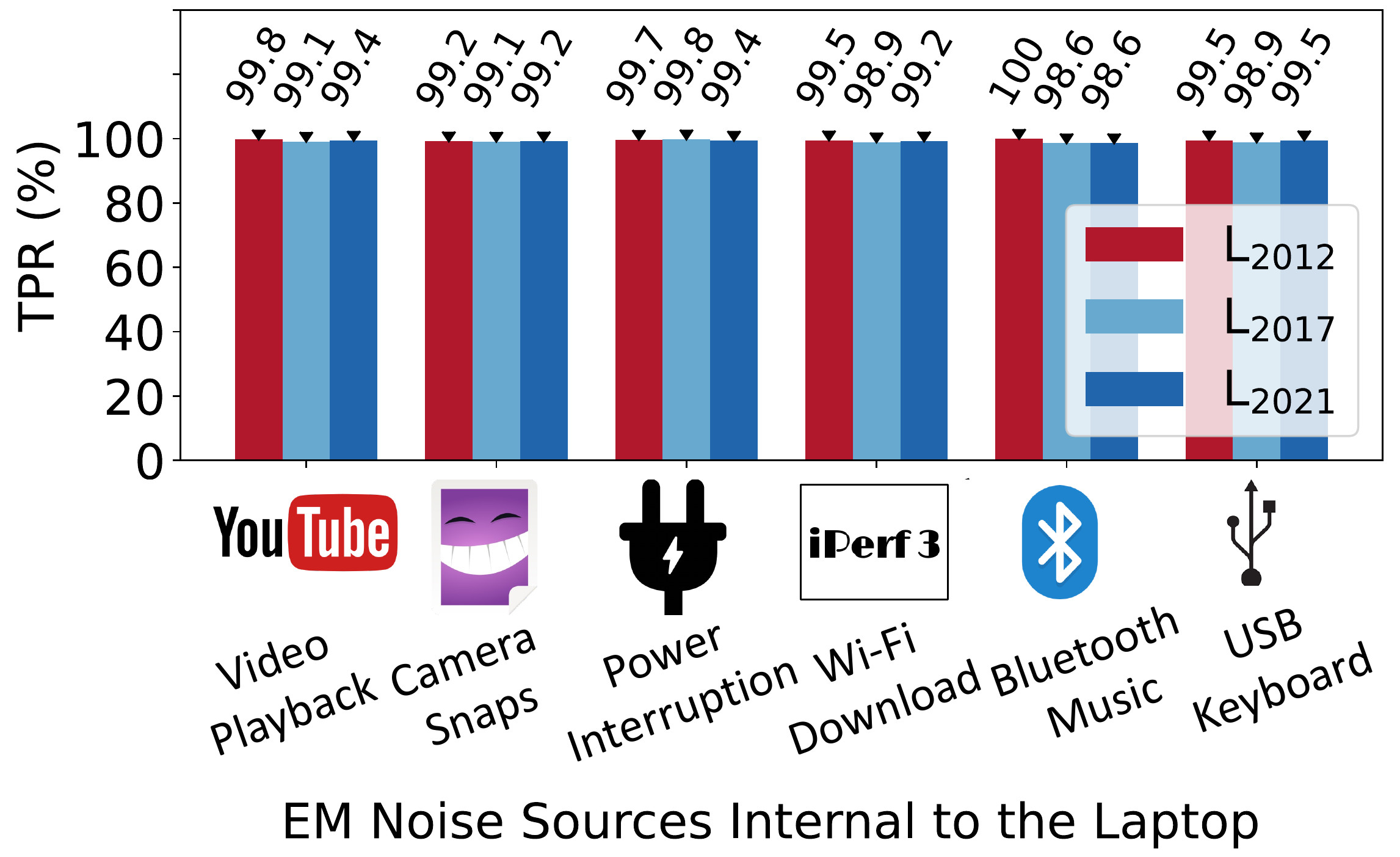}
    \caption{\revision{Figure depicts the effect on TPR due to electromagnetic noise sources internal to the laptop.}} 
    \label{fig:eval-internal-em-noise}
\end{figure}

\subsubsection{\revision{Effect of Internal Electromagnetic Noise}}
\label{sec:eval-internal-em-noise}
\revision{We evaluate the effect of electromagnetic noise arising from within the laptop, e.g., due to screen, camera and radio communication. As depicted in Figure~\ref{fig:eval-internal-em-noise}, we evaluate \name's performance when the \mic is \textit{on} in the background, along with the following six sources of EM interferences -- (1) \textit{Video Playback}: fluctuations in screen content due to high bit-rate video playback, (2) \textit{Camera Snaps}: photo capture (once every five seconds) from a camera application, (3) \textit{Power Interruption}: disruption in power (once every five seconds) due to plugging-in and plugging-out of the laptop charging cable, (4) \textit{Wi-Fi Download}: data download over Wi-Fi at 100 Mbps using iPerf3, (5) \textit{Bluetooth Music}: music playback via Bluetooth; and (6) \textit{USB Keyboard}: serial communication via USB to capture keyboard input. As depicted, we obtain a TPR above 98\% for all the three laptops across all scenarios. This good performance can be attributed to be the spatial specificity of the near-field probes to capture EM leakage in a highly localized region (i.e., within a few centimeters). }

\begin{figure}[!t]
    \centering
    \includegraphics[width=0.8\linewidth]{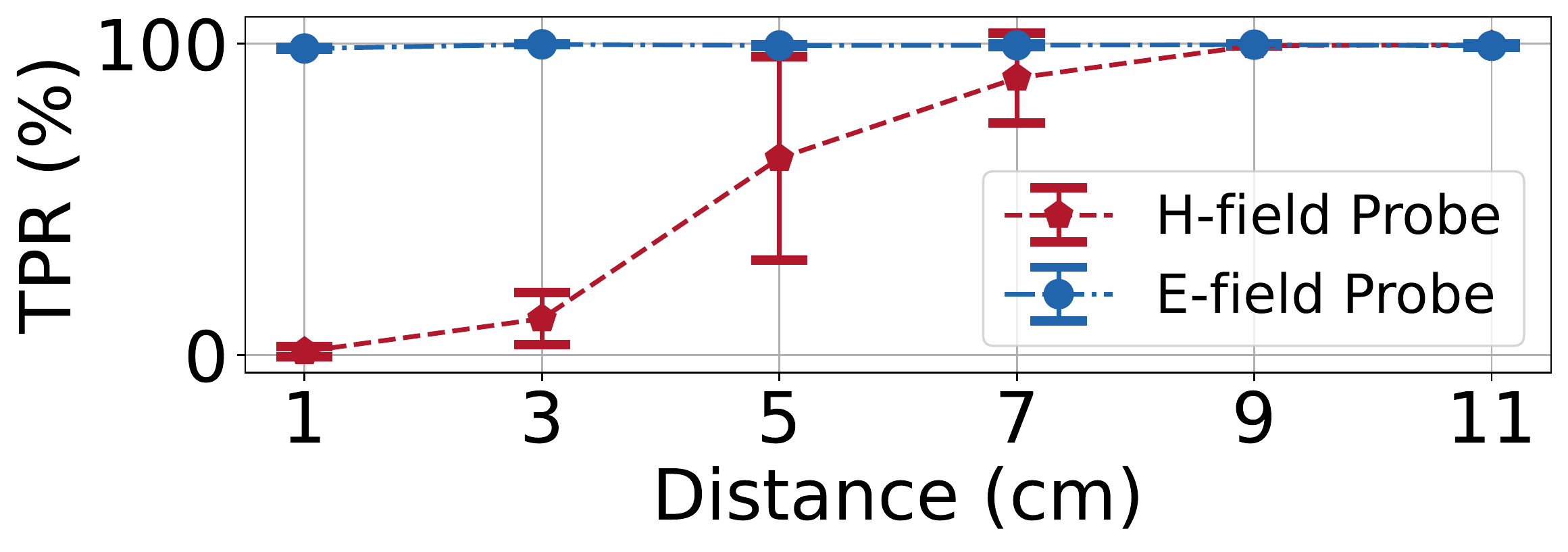}
    \caption{Figure depicts the mean (denoted by dots) and standard deviation (denoted by error bars) in TPR for EM signals captured from the near-field probes in the presence of EM disturbance from RFID readers.} 
    \label{fig:eval-noise}
\end{figure}
%
\begin{figure}[!t]
    \centering
    \includegraphics[width=0.9\linewidth]{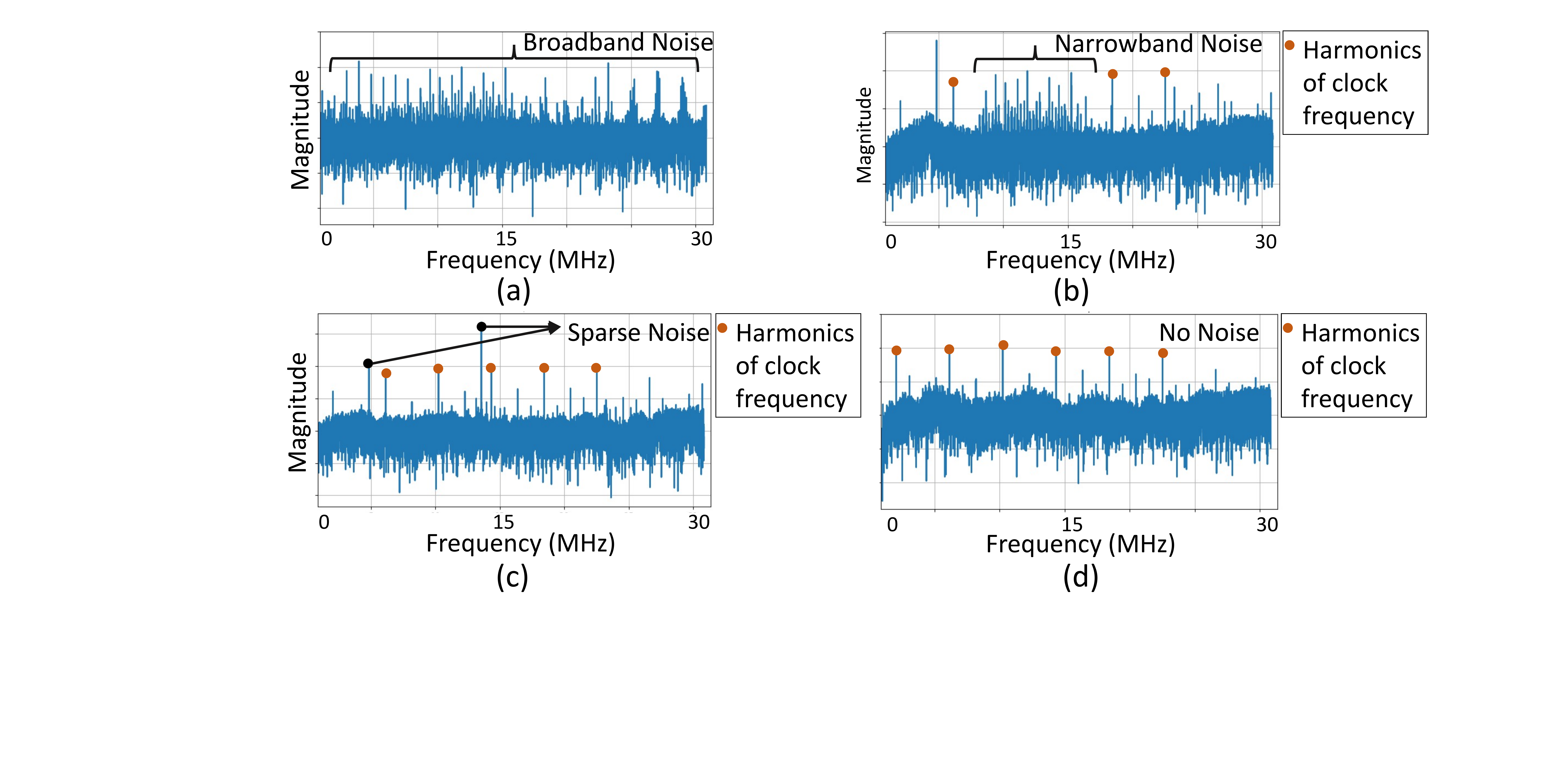}
    \caption{Figure depicts the EM leakage spectrum obtained when the RFID reader is (a) 1 cm, (b) 5 cm, and (c) 9 cm away from the H-field probe, while (d) depicts a case without an RFID reader (i.e., no noise). When the reader is 1 cm away, EM noise overshadows the \mic clock signals, while the noise drops considerably as the distance increases to 9 cm.} 
    \label{fig:eval-noise-spectrum}
\end{figure}

%
\subsubsection{Effect of External Electromagnetic Noise}
\label{sec:eval-em-noise}
Recall that the \mic clock frequencies and their harmonics are in the lower MHz range, i.e., from $1-30$ MHz. Hence, we evaluate \name in the presence of EM noise within our capture range, particularly from a radio frequency identification (RFID) reader, RFID-RC522, with a center frequency of 13.56 MHz. We test the effect on \name's TPR by varying the distance of the reader from the near-field probes, i.e., the E-field probe or the H-field probe (with 5 mm loop diameter), placed on the laptops. For this experiment, we consider three laptops, namely Lenovo Thinkpad T430U, Dell Latitude E5570, and Lenovo X1 Extreme Gen 3, that are capable of detecting \mic clock frequencies with both the above mentioned probes. 

As depicted in Figure~\ref{fig:eval-noise}, we observe that the E-field probe remains unaffected in the presence of the RFID reader, by achieving an average TPR of 98.5\% (across the three laptops) even at the closest distance of 1 cm. This is because the RFID readers create a magnetic field in the near-field region, and hence not influencing the E-field probe. On the other hand, we observe that the H-field probe is severely affected at close distances, with \name achieving an average TPR of 1.1\% at a distance of 1 cm. However, we observe that the TPR increases with distance, hence at a distance of 9 cm, the average TPR increases to a high value of 99.43\%. 
Figure~\ref{fig:eval-noise-spectrum} depicts the frequency spectrum of EM leakage for one of the laptops (i.e., Lenovo Thinkpad T430U) with the reader placed at distances of 1 cm,  5 cm, and 9 cm from the H-field probe. We observe that at the closest distance (i.e., 1 cm), the EM noise is broadband, i.e., covers a wide frequency band, and hence completely masks the underlying clock signals. However, as the distance increases, the frequency range of the noise decreases, leading to a more accurate detection of the harmonics of the \mic clock frequency. 

\begin{figure}[!t]
    \centering
    \includegraphics[width=0.8\linewidth]{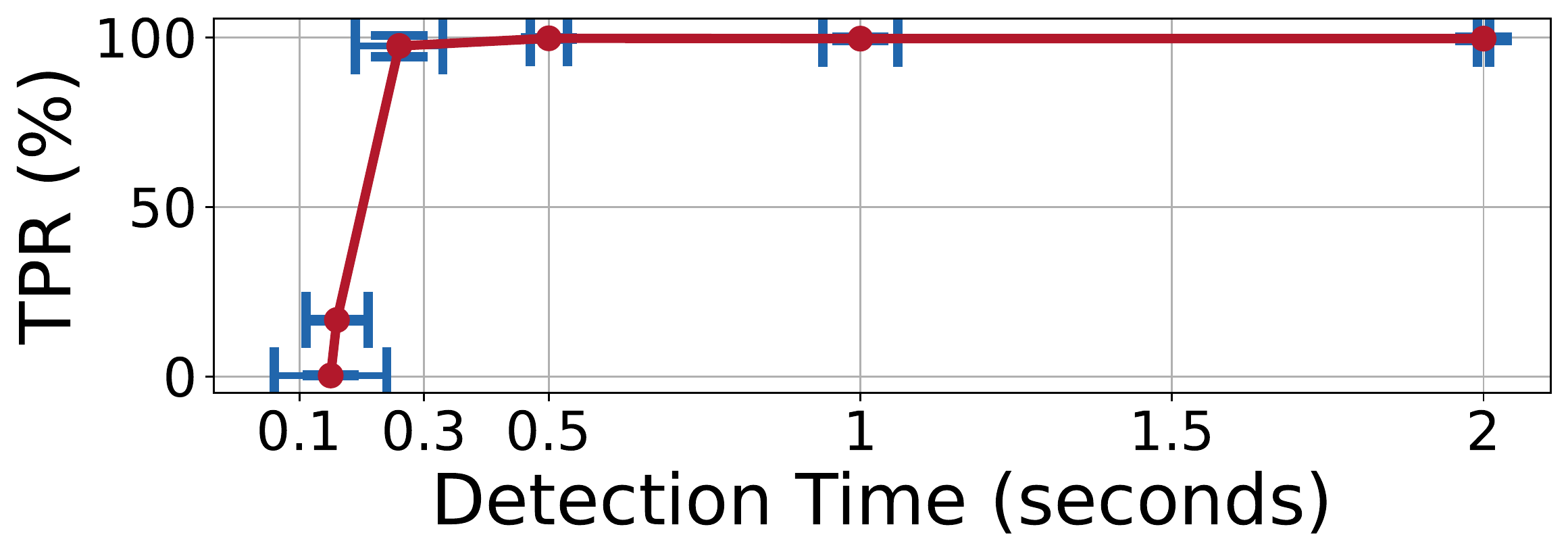}
    \caption{Figure depicts the effect of different detection times on \name's performance.} 
    \label{fig:eval-real-time}
\end{figure}
%
\subsubsection{Real-time Performance}
\label{sec:eval-real-time}
Recall that in the evaluations presented so far, we conduct \name's detection in an offline manner, i.e., we compute the clock signals present in the trace, \textit{after} all the EM traces are collected. 
In this evaluation, we test the practicality of \name by performing the detection in real-time. 
In particular, we compute the TPR by varying the detection time (i.e., time taken to output a prediction of \mic status) from $0.15 - 2$ seconds. We vary this indirectly by varying the rate at which we read from the SDR. Furthermore, we report the \textit{average} detection time as its value depends both on the frequency-switching rate of the SDR hardware as well as \name's computation time, both of which may change. 

As depicted in Figure~\ref{fig:eval-real-time}, \name achieves a high TPR of 97.5\% and 99.7\%, for average detection times of 0.26 s and 0.5 s, respectively, demonstrating the feasibility of \name 
as a real-time \mic status indicator. For average detection time lower than 0.26 s, the increase in data read-rate results in significant data overflows from the SDR, and hence results in reduction in the TPR to as low as 0.27\%, for an average detection time of 0.15 seconds.

\begin{figure}[!t]
    \centering
    \includegraphics[width=0.8\linewidth]{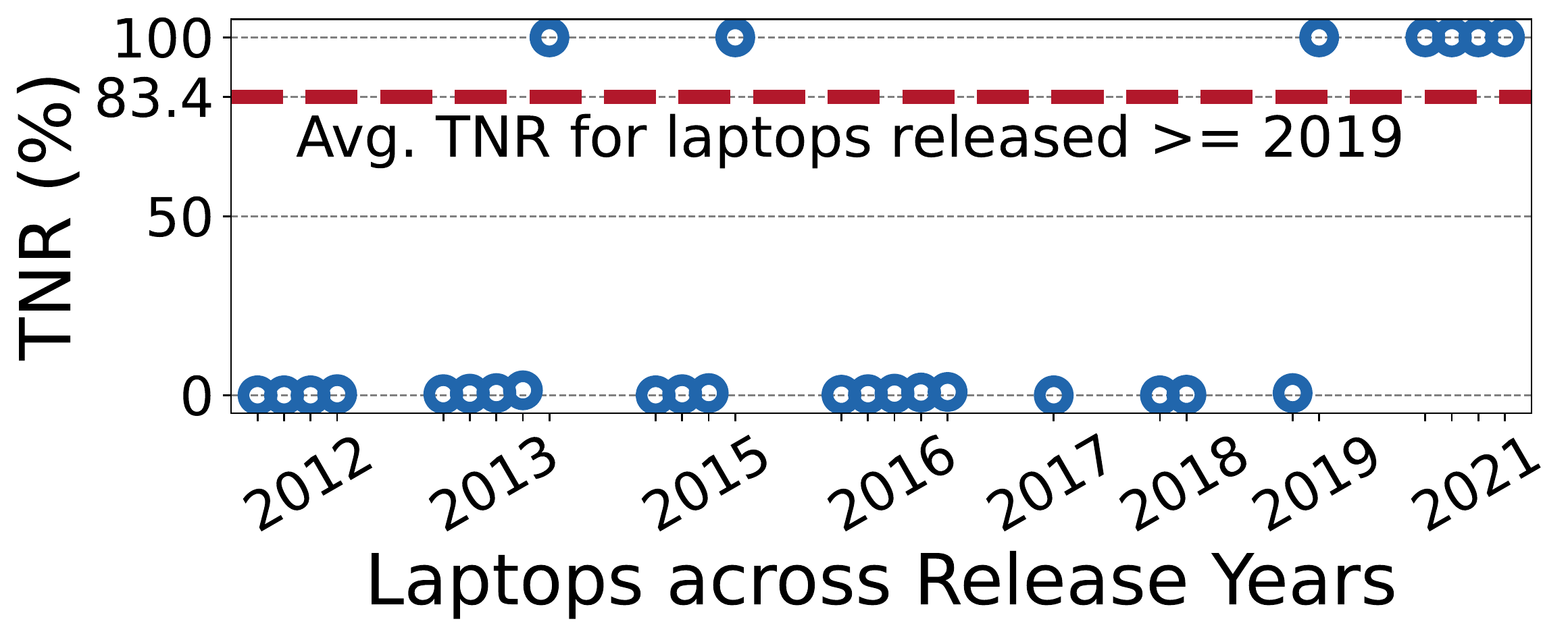}
    \caption{Figure depicts the TNR while accessing speakers for the 27 different laptop models, sorted by release year.} 
    \label{fig:eval-speaker-access}
\end{figure}
%
\subsubsection{Influence due to Speaker Access}
\label{sec:eval-speaker}
We also test for potential false triggers that may result in \mic clock frequency detection when the speaker is \textit{on}. This is because the \mic's ADC and speaker's DAC clock lines may be shared, especially if they are both controlled by the same audio codec IC. 
As this property of sharing clock lines is hardware-dependent, we perform this evaluation on all the 27 laptops.
In Figure~\ref{fig:eval-speaker-access}, we depict the \textit{true negative rate (TNR)} of all laptops, sorted chronologically by release year. We observe that in 20 out of the 27 laptops, access to speaker also triggers the same clock frequency, hence resulting in a low TNR of 26.2\% on average across all laptops. 
However, we notice a significant increase in TNR, to an average of 83.4\%, for all laptops released on or after 2019 (which includes 5 Lenovo and 1 HP laptops). This increase in TNR over the last three years seems promising, hence we believe \name has increased utility for laptops of the upcoming years.


\section{Discussion}
\label{sec:discussion}
We present important points related to \name's detection. 

\revision{\noindent\textbf{\textit{\Mic Status Detection on Non-Laptop Devices.}} 
We evaluate \name on 40 non-laptop devices, including smartphones, tablets, smart speakers and USB web-cameras. Detailed findings are reported in Appendix~\ref{app:other-devices} and Table~\ref{tbl:eval-other-devices}. 
To summarize, we successfully detect \mic clock frequency in 21 out of 40 devices. Of the successful devices, we observe an average TPR and TNR of 86.2\% ($\sigma=22.5\%$) and 100\% ($\sigma=0\%$), respectively. We note three key reasons for \name's lower detection performance on non-laptop devices:  
\begin{itemize}[leftmargin=*]
    \item \textbf{\textit{Analog vs. Digital \Mics}}: Some smartphone models contain analog \mics instead of digital \mics. 
    We believe there are several reasons for future devices to transition to digital \mics:
    Digital \mics -- (1) host an ADC, hence require fewer components to function, (2) are highly integrable into systems only containing general purpose ICs as they output digital data; and (3) are immune to EM interference compared to analog \mics, hence robust to noise. Finally, (4) digital \mics are known to be easier to design~\cite{mems-mic-2, mems-mic-3}.
    \item \textbf{\textit{Devices without Power Constraints}}: Voice-enabled smart speaker devices (including Google Home and Echo Dot) do not have any power constraints as they are always plugged-in, and may \textit{not} cut-off the clock frequency even when not recording. We observe such cases in four out of eight tested smart speakers. 
    \item \textbf{\textit{Compact Form-Factor}}: Devices with compact form-factors, e.g., smartphones enclose shorter cables (compared to laptops), and likely cause reduced EM leakage in lower radio frequencies~\cite{noise-antenna-length-murata}. 
\end{itemize}}
\noindent\textbf{\textit{Miniaturizing \name's Form-Factor.}} 
Recall from Figure~\ref{fig:prototype} that \name's current prototype consists of a variety of components stacked to the side of the laptop, while our vision is a device with a small USB drive form-factor that can be placed in contact with the laptop's exterior (Figure~\ref{fig:intro-fig}). 
One approach to reduce overall setup size is to leverage SDRs with smaller dimensions. Hence, we evaluate \name with different SDRs such as AirSpy HF+ (with small form-factor -- $45\times60\times10$ mm), and achieve high \textit{TPR} above 98\% (refer to Appendix~\ref{app:eval-sdr}). 
Another approach would be to redesign the whole setup into a single printed circuit board, consisting of the RF amplifier, a high sampling ADC (50-60 Msps), as well as the controller IC which runs \name's logic~\cite{ADC-chip-1,ADC-chip-2,ADC-chip-3,ADC-chip-4}.

\noindent\textbf{\textit{Absence of Leakage due to Clock Signals.}}  
\name's technique relies on the EM leakage from clock signals due to imperfection in hardware design including impedance mismatch at connectors, cables. Hardware designers are constantly improving the emissions from clock signals in their circuits, by incorporating techniques such as differential signaling, spread spectrum clocking, and reduction in trace length, in addition to physical methods such as shielding with metal~\cite{ti-ssc, microsemi-ssc, murata-shields, ti-emi}. However, none of these approaches are foolproof, as they can only \textit{reduce} the amount of leakage. As an example, metal shields around cables typically have slits to serve as heat vents, which can in-turn radiate EM signals in certain frequency ranges, subject to the dimensions of the slit.


\section{Related Work}
\label{sec:related-work}
We now present closely related work with \name. 

\noindent\textit{\textbf{{Eavesdropping Detection.}}}
Researchers have proposed hardware and software-based approaches to detect \mic eavesdropping~\cite{singh2021always, mitev2020leakypick, valeros2017spy, mirzamohammadi2017ditio, farley2010roving, sharma2022lumos}. 
One of the works utilizes SDRs to detect audio bugs in the environment based on their wireless transmissions~\cite{valeros2017spy}. However, none of these approaches apply to detect eavesdropping \mics in laptops. One representative work amongst the software-based approaches proposes a system for trustworthy \mic-usage notification 
by inserting run-time checks in the kernel/hypervisor~\cite{mirzamohammadi2018viola}. However, unlike these approaches, ~\name is resistant to kernel/hypervisor compromise, and its detection can easily be extended to work on devices with different specifications, e.g., different OSes.

\noindent\textit{\textbf{{Acoustic Jamming.}}}
One line of work explores generation of audio jamming signals -- both audible and inaudible, in order to prevent \mics in commodity devices 
from capturing meaningful audio~\cite{roy2017backdoor, roy2018inaudible, chen2020wearable, li2020patronus, zhu2021secure, liu2021defending}. In particular, one of the works engineered an ultrasound array in a wearable bracelet form-factor that produced inaudible jamming signals to prevent any attacker device from recording~\cite{chen2020wearable}. Our work is complementary to these works in that ~\name \textit{detects} eavesdropping \mics, while they disable them. 

\noindent\textit{\textbf{{Electromagnetic Side Channels.}}} 
There are several attacks leveraging electromagnetic leakage signals to infer cryptographic keys, screen content, passcodes, USB data, neural network architecture, and even capture audio~\cite{liu2020screen, maia2021can, jin2021periscope, camurati2018screaming, su2017usb, choi2020tempest}. However, unlike all the above, our work leverages the leaked EM signals for \textit{defense}, rather than attack. 
One particular work 
utilises leaked electromagnetic signals from local oscillators to identify wireless eavesdroppers~\cite{chaman2018ghostbuster}. However, unlike this work that detects Wi-Fi receivers, our approach detects audio receivers, i.e., \mics.


\section{Conclusion}
We present \name, a novel laptop \mic \textit{on/off} status detection, based on EM leakage of clock signals. 
We design and implement \name, as well as perform 
real-world evaluation on 30 popular laptops and observe \mic detection in 27 laptops. 
Through this work, we explore a novel direction of utilizing EM 
side-channel information as part of a defense, 
rather than an attack. As part of future work, we hope 
to utilize \name to identify access to 
other sensors including cameras and inertial measurement unit (IMU) sensors. 

\section{Acknowledgements}
We thank Wang Gucheng, Nitya Lakshmanan, Siddharth Rupavatharam and Niel Warren for valuable discussions and/or feedback on our paper. 
This work is supported by the Singapore Ministry of Education Academic Research Fund (R-252-000-B48-114), the Yonsei University Research Fund (2021-22-0337), the Institute of Information and Communications Technology Planning and Evaluation (IITP-2022-0-00420) grant funded by Ministry of Science and ICT (MSIT) in Korea, and the Google PhD Fellowship 2021. 

\bibliographystyle{ACM-Reference-Format}
\bibliography{paper}

\balance

\appendix

\section{Additional Evaluation}
\label{app:additional-eval}
In this section, we elaborate on the choice of parameter values in our evaluation, as well as provide some additional results.

\subsection{Parameter Selection}
\label{app:module-eval}

\begin{figure}[!t]
    \centering
    \includegraphics[width=0.8\linewidth]{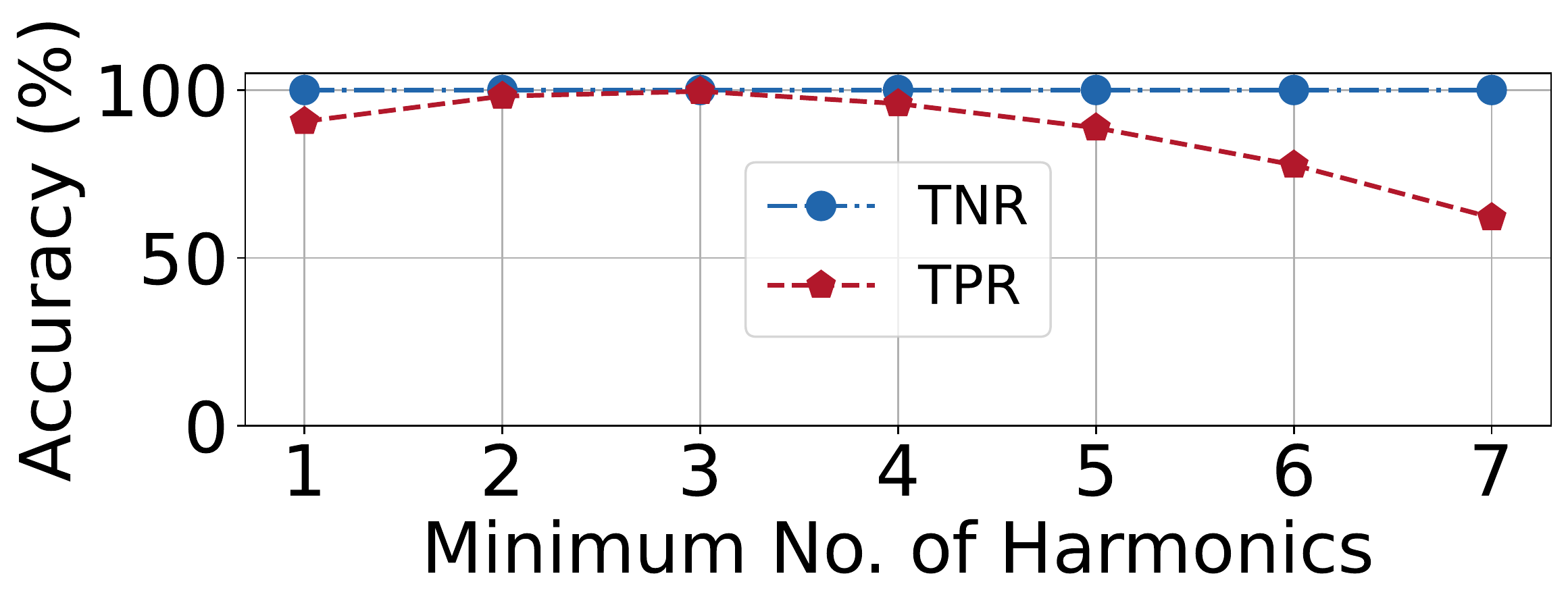}
    \caption{Figure depicts the TPR and FPR for varying values of the minimum number of harmonics ($\theta_h$) parameter.} 
    \label{fig:eval-module-eval}
\end{figure}

We present results to justify the value of parameter, $\theta_h$, used in our experimental evaluation section (\S\ref{sec:evaluation}). 
Figure~\ref{fig:eval-module-eval} depicts the overall TPR and TNR obtained on the $27$ laptops (see \S\ref{sec:eval-overall}) for different values of the parameter, $\theta_h$, ranging from $1-7$.  
As illustrated in the figure, the TNR is almost 100\% (i.e., above 99.99\%) irrespective of the choice of the number of harmonics, highlighting \name's robustness to false positives. However, the TPR varies with $\theta_h$, where we achieve the best TPR of 99.6\% (i.e., $\frac{17480}{17550}$) at the value, $\theta_h=3$, and the second-best result of 98.2\% (i.e., $\frac{17229}{17550}$), for $\theta_h=2$. Hence, we set the value of minimum number of harmonics to three, i.e., $\theta_h=3$. 

\subsection{\name's Performance in the Presence of Audio of Varying Sound Levels}
\label{app:eval-sound-levels}

\begin{figure}[!t]
    \centering
    \includegraphics[width=0.8\linewidth]{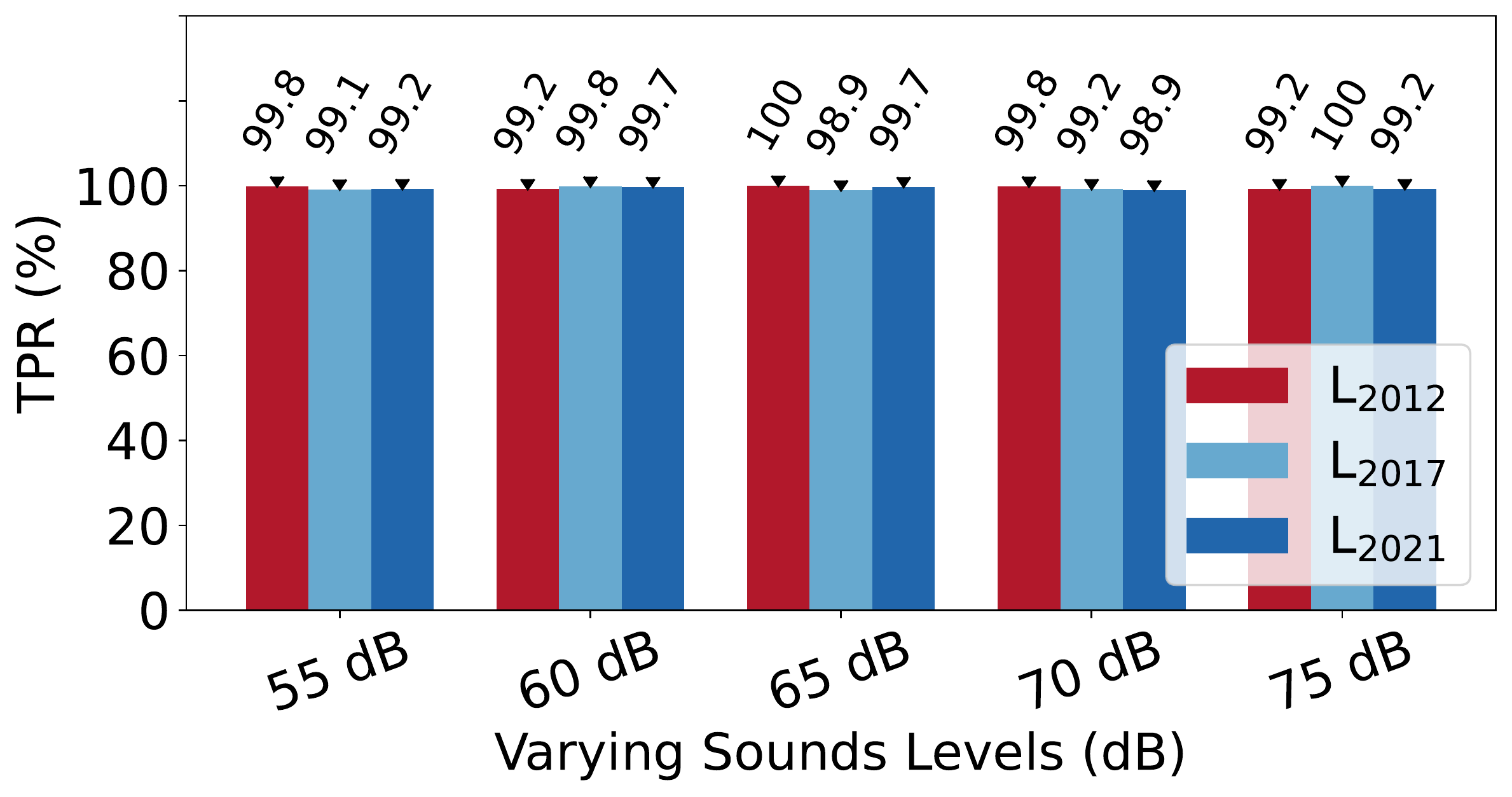}
    \caption{\revision{Figure depicts the effect on varying sound levels on ~\name's TPR performance.}} 
    \label{fig:eval-diff-sound-levels}
\end{figure}

In Section~\ref{sec:eval-mic-app}, we evaluate \name's performance in the presence of multiple \mic-based applications. However, we do not control the incoming audio sound level which may additionally influence the EM leakage level. Hence, as a follow-up, we now evaluate the effect of the ambient sound levels 
as captured by the laptop \mic on \name's performance. Specifically, we perform experiments on the laptops, \lold, \lmid, and \lnew
~(defined in Section \ref{sec:eval-differing-cond}). 
In order to produce audio with varying sound levels, we utilize an external speaker (Adam Audio A3X studio monitor~\cite{adam-studio-monitor}), which we place about $15$ centimeters away from the laptop, and play audio signals with varying loudness levels, ranging from $55 - 75$ dB and record sound from the laptop. 
In particular, we play white noise, consisting of uniform energies in all frequency bins from 0-24 kHz. As depicted in Figure~\ref{fig:eval-diff-sound-levels}, we obtain a TPR above 98\% across all noise levels and laptops. Hence, we find that the varying ambient sound levels cast negligible influence on \name's performance.

%
\section{Overall Laptop Results}
\label{app:laptop-results}
%
\begin{table*}[]
\begin{tabular}{|l|l|c|c|c|c|c|c|c|c|}
\hline
S.No & Device Model & \multicolumn{1}{l|}{Year} & \multicolumn{1}{l|}{\begin{tabular}[c]{@{}l@{}}Freq\\ \fclk\\ (kHz)\end{tabular}} & \begin{tabular}[c]{@{}l@{}}Best\\ Probe\end{tabular} & \multicolumn{1}{l|}{\begin{tabular}[c]{@{}l@{}}Best True\\ Positive Score\\ (/650)\end{tabular}} & \multicolumn{1}{l|}{\begin{tabular}[c]{@{}l@{}}Leak\\ Level\\ (dB)\end{tabular}} & O/E & \multicolumn{1}{l|}{\begin{tabular}[c]{@{}l@{}}E-field True\\ Positive Score\\ (/650)\end{tabular}} & \multicolumn{1}{l|}{\begin{tabular}[c]{@{}l@{}}H-field True\\ Positive Score\\ (/650)\end{tabular}} \\ \hline
1 & ASUS Strix GL502VT & 2016 & 2048 & h-field & 646 & 27 & O & 531 & 646 \\ \hline
2 & Asus X450v & 2013 & 2048 & e-field & 642 & 31 & O & 642 & 641 \\ \hline
3 & Dell Inspiron 13 7359 & 2015 & 3072 & h-field & 649 & 36 & O & 648 & 649 \\ \hline
4 & Dell Inspiron 5459 & 2015 & 2048 & e-field & 650 & 25 & O & 650 & 648 \\ \hline
5 & Dell Inspiron 7572 & 2018 & 2048 & e-field & 650 & 27 & O & 650 & 641 \\ \hline
6 & Dell Latitude E5570 & 2016 & 2048 & e-field & 642 & 30 & O & 642 & 641 \\ \hline
7 & Dell Latitude E7450 & 2015 & 2048 & e-field & 650 & 26 & O & 650 & 649 \\ \hline
8 & Dell XPS L321x & 2012 & 2048 & h-field & 650 & 35 & O & 647 & 650 \\ \hline
9 & Fujitsu Lifebook SH772 & 2013 & 2048 & e-field & 650 & 35 & O & 650 & 648 \\ \hline
10 & HP Envy 13 & 2021 & 3072 & h-field & 648 & 23 & O & -- & 648 \\ \hline
11 & HP Probook 440 G1 & 2013 & 2352 & e-field & 650 & 21 & O & 650 & 649 \\ \hline
12 & HP Zbook Studio G5 & 2018 & 3072 & h-field & 648 & 28 & O,E & 646 & 648 \\ \hline
13 & Lenovo P14s G1 & 2021 & 2400 & h-field & 650 & 24 & O & 649 & 650 \\ \hline
14 & Lenovo T430U & 2012 & 2048 & h-field & 648 & 34 & O & 646 & 648 \\ \hline
15 & Lenovo T460s & 2016 & 2048 & h-field & 645 & 25 & O & -- & 645 \\ \hline
16 & Lenovo T470S & 2017 & 3072 & h-field & 650 & 25 & O & 647 & 650 \\ \hline
17 & Lenovo T590 & 2019 & 2048 & h-field & 643 & 23 & O & 633 & 643 \\ \hline
18 & Lenovo X1 Extreme G3 & 2021 & 2400 & h-field & 649 & 24 & O & 644 & 649 \\ \hline
19 & Lenovo X13 Gen 2 & 2021 & 2400 & h-field & 646 & 25 & O & -- & 646 \\ \hline
20 & Lenovo X1 Carbon G7 & 2019 & 2400 & h-field & 646 & 22 & O,E & -- & 646 \\ \hline
21 & Lenovo X230 & 2012 & 2048 & h-field & 650 & 31 & O & -- & 650 \\ \hline
22 & Lenovo X250 & 2015 & 2048 & h-field & 641 & 24 & O & -- & 641 \\ \hline
23 & Lenovo X260 & 2016 & 2048 & h-field & 649 & 21 & O & -- & 649 \\ \hline
24 & Razer RZ09-0102 & 2013 & 2048 & h-field & 650 & 26 & O & 645 & 650 \\ \hline
25 & Samsung Chronos NP770Z5E & 2013 & 2048 & e-field & 650 & 28 & O & 650 & 646 \\ \hline
26 & Terrans Force T5 & 2016 & 2048 & h-field & 649 & 30 & O & 648 & 649 \\ \hline
27 & Toshiba Portege Z930 & 2012 & 6144 & h-field & 639 & 23 & O,E & -- & 639 \\ \hline
{\color{red} 28} & {\color{red} Macbook Pro 2017 13"} & {\color{red} 2017} & {\color{red} 2823} & {\color{red} h-field} & {\color{red} 642} & {\color{red} 18} & {\color{red} O} & {\color{red} --} & {\color{red} 642} \\ \hline
{\color{red} 29} & {\color{red} Macbook Pro 2014 15"} & {\color{red} 2014} & {\color{red} --} & {\color{red} --} & {\color{red} --} & {\color{red} --} & {\color{red} --} & {\color{red} --} & {\color{red} --} \\ \hline
{\color{red} 30} & {\color{red} Macbook Pro 2019 16"} & {\color{red} 2019} & {\color{red} --} & {\color{red} --} & {\color{red} --} & {\color{red} --} & {\color{red} --} & {\color{red} --} & {\color{red} --} \\ \hline
\end{tabular}
\caption{Table depicts the results of the 30 tested laptops, along with their release year, near-field probe that resulted in best results (i.e., higher true positive score), best true positive score, leakage amplitude, harmonics detected (O:odd, E:even), true positive scores of E-field probe and H-field probe (-- indicates the microphone clock frequency, ~\fclk, not detected).}
\label{tbl:laptop-overall-appendix}
\end{table*}

This section presents the detailed results of the overall evaluation of 30 tested laptops. Table~\ref{tbl:laptop-overall-appendix} depicts the results, including 27 successful laptops, one confounding case (Macbook Pro 2017), as well as the two unsuccessful cases. We tabulate the results obtained with each of the near-field probes, as well as the best results (which is reported in \S\ref{sec:eval-overall}). We observe that the H-field probe successfully detects the microphone clock frequency, ~\fclk, across all laptops, while E-field probe successfully detects 19 out of 27 laptops. For the H-field probe, we leverage loops with diameter from 5 mm up to 20 mm, where larger loops has higher sensitivity and hence captures weaker signals. Furthermore, we observe that most laptops, (i.e., 24/27) have only odd harmonics of the microphone clock frequency observed {\em prominently}. By prominently, we mean more than 80\% of all detected harmonics are odd. However, in the remaining three cases, both odd and even harmonics occur, which could be due to a deviation of clock duty cycle from 50\% (\S\ref{sec:bg-clock}).  

\begin{table}[]
\begin{tabular}{|l|l|l||l|l|l|}
\hline
\begin{tabular}[c]{@{}l@{}}MacBook\\ Model\end{tabular} &
  \begin{tabular}[c]{@{}l@{}}Freq\\ \fclk\\ (kHz)\end{tabular} &
  \begin{tabular}[c]{@{}l@{}}Unique\\ Clk?\end{tabular} &
  \begin{tabular}[c]{@{}l@{}}MacBook\\ Model\end{tabular} &
  \begin{tabular}[c]{@{}l@{}}Freq\\ \fclk\\ (kHz)\end{tabular} &
  \begin{tabular}[c]{@{}l@{}}Unique\\ Clk?\end{tabular} \\ \hline
Pro 14 & 1384 & \cmark & Pro 19 & -- & {\color{red} \xmark} \\ \hline
Pro 15 & -- & {\color{red} \xmark} & Pro 20 & -- & {\color{red} \xmark} \\ \hline
Pro 16 & -- & {\color{red} \xmark} & Air 13 & 1613 & \cmark  \\ \hline
Pro 17 & 2832 & \cmark & Air 15 & -- & {\color{red} \xmark} \\ \hline
Pro 18 & -- & {\color{red} \xmark} & Air 20 & 2314 & \cmark  \\ \hline
\end{tabular}
\caption{\revision{Table represents our analysis on ten Macbook models, along with the detected \mic clock frequency, \fclk~(if any).}}
\label{tbl:eval-macbook}
\end{table}
\revision{\subsection{Macbook Performance}
\label{app:sec-macbook}

Recall from Section~\ref{sec:eval-overall} that \name fails to detect the \mic clock frequency, \fclk, in three Macbook models. Hence, we perform comprehensive evaluation on ten different Macbook models using a high-performance setup. The setup uses a Sonoma Broadband RF Amplifier, operating between 10 kHz - 2.5 GHz, with a 38 dB gain at 1 GHz, 
as well as an Anritsu MS2668C spectrum analyzer~\cite{sonoma-rf-amp, anritsu-spectrum-analyzer}. On capturing the EM signals while recording using the \textit{Voice Memos} application, we observe a unique clock frequency, in four out of the ten models (Table~\ref{tbl:eval-macbook}). However, these frequencies do not correspond to the commonly found \mic clock frequencies in other devices (which are typically multiples of an audio sampling rate), and seem to vary across Macbook models. 
We conjecture two possible reasons for \name's poor performance on Macbooks -- (1) The clock signal of the \mic itself is well shielded by the Macbook, and there is no significant signal leakage due to the \mic in a majority of tested models (i.e., six out of ten). (2) Macbooks utilize custom-designed \mics and/or codec that communicate over non-standard protocols. Hence, clock frequencies utilized deviate from the commonly used frequencies.}

\begin{table*}[]
\begin{tabular}{|lrll|lrll|lrll|}
\hline
\multicolumn{1}{|l|}{Device Model} & \multicolumn{1}{l|}{\begin{tabular}[c]{@{}l@{}}\fclk\\ (kHz)\end{tabular}} & \multicolumn{1}{c|}{\begin{tabular}[c]{@{}c@{}}Unique\\Clk?\end{tabular}} & A/D & \multicolumn{1}{l|}{Device Model} & \multicolumn{1}{l|}{\begin{tabular}[c]{@{}l@{}}\fclk\\ (kHz)\end{tabular}} & \multicolumn{1}{c|}{\begin{tabular}[c]{@{}c@{}}Unique\\Clk?\end{tabular}} & A/D & \multicolumn{1}{l|}{Device Model} & \multicolumn{1}{l|}{\begin{tabular}[c]{@{}l@{}}\fclk\\ (kHz)\end{tabular}} & \multicolumn{1}{c|}{\begin{tabular}[c]{@{}c@{}}Unique\\Clk?\end{tabular}} & A/D \\ \hline
\multicolumn{4}{|c|}{\textbf{Smartphones}} &
  \multicolumn{1}{l|}{Samsung A01 Core} &
  \multicolumn{1}{r|}{--} &
  \multicolumn{1}{l|}{\color{red}\xmark} &
  A &
  \multicolumn{1}{l|}{Valore VM64} &
  \multicolumn{1}{r|}{3072} &
  \multicolumn{1}{l|}{{\cmark}} &
  D \\ \hline
\multicolumn{1}{|l|}{Pixel 3XL} &
  \multicolumn{1}{r|}{1200} &
  \multicolumn{1}{l|}{{\cmark}} &
  U &
  \multicolumn{1}{l|}{Samsung A52s} &
  \multicolumn{1}{r|}{--} &
  \multicolumn{1}{l|}{\color{red}\xmark} &
  A &
  \multicolumn{1}{l|}{Logitech C505} &
  \multicolumn{1}{r|}{--} &
  \multicolumn{1}{l|}{\color{red}\xmark} &
  A \\ \hline
\multicolumn{1}{|l|}{\revision{Pixel 5a}} &
  \multicolumn{1}{r|}{2048} &
  \multicolumn{1}{l|}{{\cmark}} &
  U &
  \multicolumn{1}{l|}{Samsung A6+} &
  \multicolumn{1}{r|}{--} &
  \multicolumn{1}{l|}{\color{red}\xmark} &
  A &
  \multicolumn{1}{l|}{Logitech C930E} &
  \multicolumn{1}{r|}{--} &
  \multicolumn{1}{l|}{\color{red}\xmark} &
  A \\ \hline
\multicolumn{1}{|l|}{\revision{Pixel 6 Pro}} &
  \multicolumn{1}{r|}{3072} &
  \multicolumn{1}{l|}{{\cmark}} &
  U &
  \multicolumn{4}{c|}{\textbf{Tablets}} &
  \multicolumn{1}{l|}{Tecgear Sentinel} &
  \multicolumn{1}{r|}{--} &
  \multicolumn{1}{l|}{\color{red}\xmark} &
  A \\ \hline
\multicolumn{1}{|l|}{Samsung A8+} &
  \multicolumn{1}{r|}{3072} &
  \multicolumn{1}{l|}{{\cmark}} &
  D &
  \multicolumn{1}{l|}{iPad Gen 3} &
  \multicolumn{1}{r|}{3000} &
  \multicolumn{1}{l|}{{\cmark}} &
  U &
  \multicolumn{1}{l|}{Xiaomi Xiaovv} &
  \multicolumn{1}{r|}{--} &
  \multicolumn{1}{l|}{\color{red}\xmark} &
  A \\ \hline
\multicolumn{1}{|l|}{Samsung A9} &
  \multicolumn{1}{r|}{2400} &
  \multicolumn{1}{l|}{{\cmark}} &
  D &
  \multicolumn{1}{l|}{Lenovo Chromebook} &
  \multicolumn{1}{r|}{3250} &
  \multicolumn{1}{l|}{{\cmark}} &
  U &
  \multicolumn{4}{c|}{\textbf{\revision{Voice-enabled Smart Speakers}}} \\ \hline
\multicolumn{1}{|l|}{Samsung Note 10+} &
  \multicolumn{1}{r|}{3072} &
  \multicolumn{1}{l|}{{\cmark}} &
  D &
  \multicolumn{1}{l|}{Surface Pro G1} &
  \multicolumn{1}{r|}{2048} &
  \multicolumn{1}{l|}{{\cmark}} &
  U &
  \multicolumn{1}{l|}{\revision{Facebook Portal Mini}} &
  \multicolumn{1}{r|}{2400} &
  \multicolumn{1}{l|}{{\cmark}} &
  U \\ \hline
\multicolumn{1}{|l|}{Samsung Note 20 U} &
  \multicolumn{1}{r|}{3072} &
  \multicolumn{1}{l|}{{\cmark}} &
  D &
  \multicolumn{1}{l|}{Surface Pro G5} &
  \multicolumn{1}{r|}{3000} &
  \multicolumn{1}{l|}{{\cmark}} &
  U &
  \multicolumn{1}{l|}{\revision{Google Nest Hub G2}} &
  \multicolumn{1}{r|}{3072} &
  \multicolumn{1}{l|}{{\cmark}} &
  U \\ \hline
  \multicolumn{1}{|l|}{Samsung S20 U} &
  \multicolumn{1}{r|}{3072} &
  \multicolumn{1}{l|}{{\cmark}} &
  D &
  \multicolumn{1}{l|}{Surface Pro G6} &
  \multicolumn{1}{r|}{3000} &
  \multicolumn{1}{l|}{{\cmark}} &
  U &
  \multicolumn{1}{l|}{\revision{Echo Show 5 G2}} &
  \multicolumn{1}{r|}{--} &
  \multicolumn{1}{l|}{\color{red}\xmark} &
  U \\ \hline

\multicolumn{1}{|l|}{Samsung S20+} &
  \multicolumn{1}{r|}{3072} &
  \multicolumn{1}{l|}{{\cmark}} &
  D &
  \multicolumn{1}{l|}{iPad Gen 8} &
  \multicolumn{1}{r|}{--} &
  \multicolumn{1}{l|}{\color{red}\xmark} &
  U &
  \multicolumn{1}{l|}{\revision{Google Home Mini}} &
  \multicolumn{1}{r|}{768} &
  \multicolumn{1}{l|}{\color{brown}$\otimes$} &
  D \\ \hline
\multicolumn{1}{|l|}{Samsung S21} &
  \multicolumn{1}{r|}{3072} &
  \multicolumn{1}{l|}{{\cmark}} &
  D &
  \multicolumn{1}{l|}{Lenovo M10} &
  \multicolumn{1}{r|}{--} &
  \multicolumn{1}{l|}{\color{red}\xmark} &
  A &
  \multicolumn{1}{l|}{\revision{Google Nest Mini}} &
  \multicolumn{1}{r|}{--} &
  \multicolumn{1}{l|}{\color{red}\xmark} &
  U \\ \hline
\multicolumn{1}{|l|}{Huawei P30 Pro} &
  \multicolumn{1}{r|}{--} &
  \multicolumn{1}{l|}{\color{red}\xmark} &
  U &
  \multicolumn{4}{c|}{\textbf{USB Webcam}} &
  \multicolumn{1}{l|}{\revision{Homepod Mini}} &
  \multicolumn{1}{r|}{2400} &
  \multicolumn{1}{l|}{\color{brown}$\otimes$} &
  U \\ \hline
\multicolumn{1}{|l|}{iPhone 8+} &
  \multicolumn{1}{r|}{--} &
  \multicolumn{1}{l|}{{\color{brown}$\otimes$}} &
  U &
  \multicolumn{1}{l|}{Asus C3} &
  \multicolumn{1}{r|}{3073} &
  \multicolumn{1}{l|}{\cmark} &
  A &
  \multicolumn{1}{l|}{\revision{Lenovo Smart Clock}} &
  \multicolumn{1}{r|}{3072} &
  \multicolumn{1}{l|}{\color{brown}$\otimes$} &
  U \\ \hline
\multicolumn{1}{|l|}{One Plus Nord 10} &
  \multicolumn{1}{r|}{--} &
  \multicolumn{1}{l|}{\color{red}\xmark} &
  U &
  \multicolumn{1}{l|}{Creative Live} &
  \multicolumn{1}{r|}{3072} &
  \multicolumn{1}{l|}{{\cmark}} &
  D &
  \multicolumn{1}{l|}{\revision{Mi Smart Clock}} &
  \multicolumn{1}{r|}{3250} &
  \multicolumn{1}{l|}{\color{brown}$\otimes$} &
  U \\ \hline
\multicolumn{1}{|l|}{One Plus Nord CE} &
  \multicolumn{1}{r|}{--} &
  \multicolumn{1}{l|}{\color{red}\xmark} &
  U &
  \multicolumn{1}{l|}{Hyso 1080p} &
  \multicolumn{1}{r|}{2048} &
  \multicolumn{1}{l|}{{\cmark}} &
  U &
  \multicolumn{1}{l|}{} &
  \multicolumn{1}{l|}{} &
  \multicolumn{1}{l|}{} &
   \\ \hline
\end{tabular}
\caption{\revision{\name's performance on different device types, including smartphones, tablets, USB web cameras and voice-enabled smart speakers, along with the value of the detected microphone clock frequency, ~\fclk, along with an indicator (i.e., \cmark vs. {\color{red}\xmark}) to denote whether this frequency is uniquely present \textit{only} when the \mic is \textit{on}. 
Finally, the A/D column depicts if the microphone is -- A: analog, D: digital, or U: unknown. {\color{brown}$\otimes$} mark indicates confounding special cases (see corresponding text for detailed explanation).}}
\label{tbl:eval-other-devices}
\end{table*}

%
\begin{figure}[!t]
     \centering
     \begin{subfigure}[b]{0.8\linewidth}
         \centering
         \includegraphics[width=\textwidth]{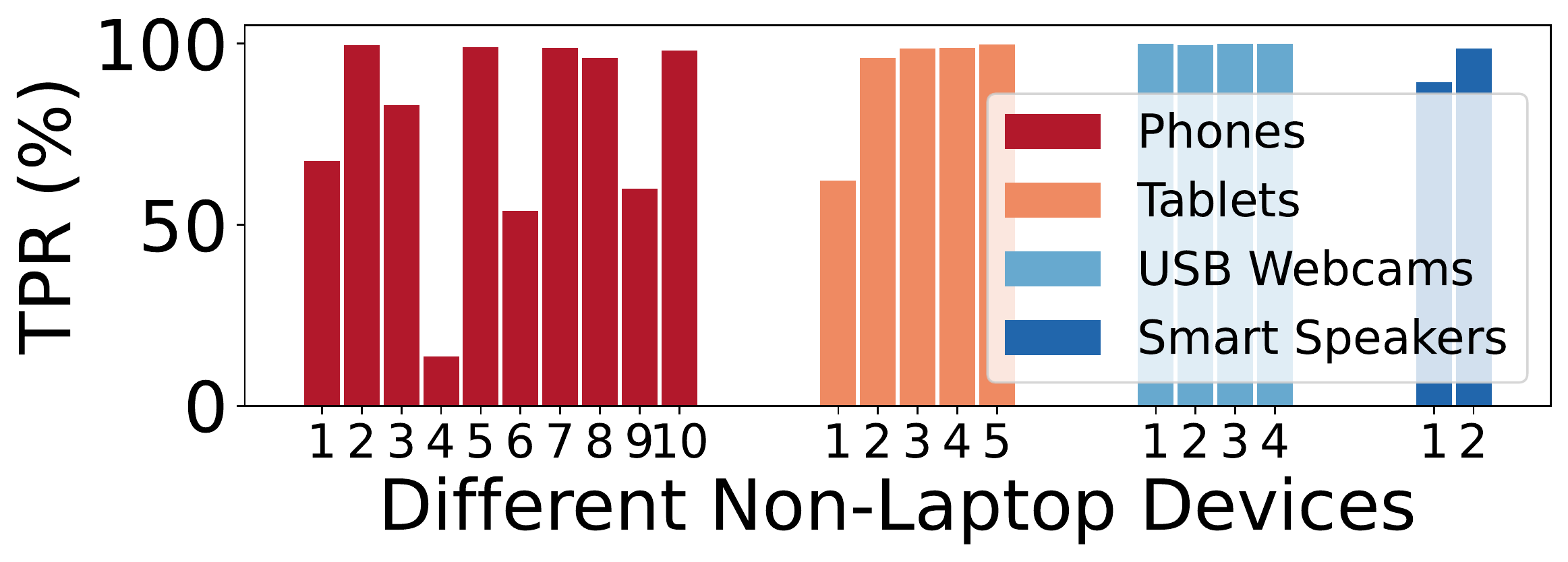}
         \caption{}
         \label{fig:eval-other-devices-tpr}
     \end{subfigure}
     \hfill
     \begin{subfigure}[b]{0.8\linewidth}
         \centering
         \includegraphics[width=\textwidth]{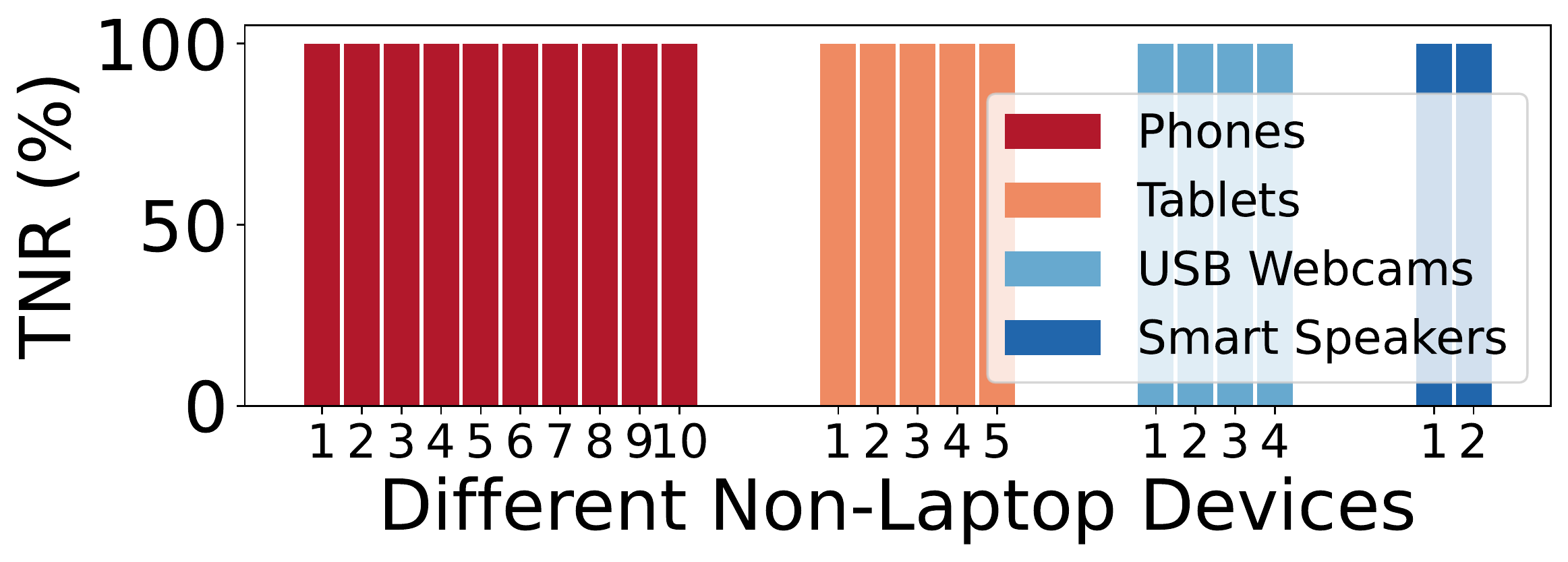}
         \caption{}
         \label{fig:eval-other-devices-tnr}
     \end{subfigure}
     \caption{\revision{Figure depicts the (a) {\em true positive rate}, and (b) {\em true negative rate} for devices of type -- smartphones, tablets, web-cameras and smart speakers.}}
     \label{fig:eval-other-devices-per-device}
\end{figure}

\section{Performance of Non-Laptop Devices}
\label{app:other-devices}
Apart from laptops, we also evaluate \name on four consumer device categories, namely smartphones, tablets, USB web-cameras and voice-enabled devices to test the generalizability of the approach. 

\subsubsection*{\textbf{Smartphones}} 

As depicted in Table~\ref{tbl:eval-other-devices}, we evaluate \name on 17 smartphones, from popular manufacturers including Samsung, Google and One Plus. We capture traces when the primary \mic (i.e., bottom \mic) is {\em on}, 
as well as when the \mic is {\em off}. The screen is kept active throughout data collection, except for phones, Pixel 3XL, Pixel 6 Pro and Samsung A8 Plus as their performance is affected by the screen's EM disturbance. 
From the table, we observe that \name successfully identifies the \mic clock frequency, ~\fclk, in 10 out of the 17 tested smartphones, including Samsung's flagship S-series phones (e.g., Galaxy S20 and S21). Furthermore, as depicted in Figure~\ref{fig:eval-other-devices-per-device}, we achieve an average {\em TPR} of 76.9\% ($\sigma = 28.3\%$), and TNR of 100\% across the 10 phones.  
Interestingly, we encounter a \textit{confounding case} ({\color{brown}$\otimes$}) with Apple iPhone 8 Plus, where we obtain inconsistent results across different audio recording applications. We conjecture that the reason is similar to the conjecture we make for Apple Macbooks (see Appendix~\ref{app:sec-macbook}). 

\subsubsection*{\textbf{Tablets}}

We evaluate seven tablets, including iPad, Chromebook and Surface Pro, by capturing traces when the microphone is {\em on}, and {\em off}, with the screen active throughout the data collection. As depicted in Table~\ref{tbl:eval-other-devices}, we identify the microphone clock frequency, ~\fclk, in 5 out of the 7 tablets. Of the two unsuccessful cases, we suspect that one of them houses an analog microphone~\cite{lenovo-m10}. Furthermore, on the five working cases, ~\name achieves a average {\em TPR} and {\em TNR} of 91.0\% ($\sigma = 16.2\%$) and 100\% respectively, with the tablet, iPad 3 alone, achieving a low {\em TPR} around 62.2\%. 

\subsubsection*{\textbf{USB Web-Cameras}}

We evaluate \name on eight USB web-cameras containing \mics, of which we successfully identify the \mic clock frequency, ~\fclk, in 4 of them. We perform device tear-downs and confirm that all the remaining four web cameras do not leak the clock frequencies as they enclose analog microphones. 
Furthermore, we encounter an interesting case where a web-cam containing analog \mic (Asus C3) also 
leaks the \mic clock frequency (\fclk~$=3.073$ MHz). We confirm that this occurs due to the leakage of the clock signal directly from the ADC chip, ES7243, itself (rather than cables and connectors), 
due to the proximity of the 
ADC chip to the plastic exterior of the webcam~\cite{adc-chip}. 
Finally, for the four successful webcams, we report a high average {\em TPR} and {\em TNR} of 99.8\% ($\sigma = 0.1\%$) and 100\% respectively (Figure~\ref{fig:eval-other-devices-per-device}).

\revision{
\subsubsection*{\textbf{Voice-enabled Smart Speakers}}

We evaluate \name on eight smart-speaker devices (Table~\ref{tbl:eval-other-devices}). As these devices are always listening for a wake-word (e.g., \textit{``Hey Google''}, \textit{``Alexa''}), we capture EM traces during normal operation (i.e., \textit{unmuted} idle mode), as well as when they are explicitly \textit{muted} (by pressing the physical mute button), to represent the \mic's \textit{on} and \textit{off} phases, respectively.  
We observe that in two out of the eight tested devices, namely Portal Mini and Nest Hub G2, we obtain unique \mic clock frequencies,~\fclk, of 2400~kHz and 3072~kHz, respectively. Furthermore, as depicted in Figure~\ref{fig:eval-other-devices-per-device}, we achieve an average TPR and TNR of 94\% ($\sigma=6.5\%$) and 100\% for the two smart speakers. 
Interestingly, we also find \textit{confounding cases} ({\color{brown}$\otimes$}) in four out of the eight devices -- Google Home Mini, Homepod Mini, Lenovo Smart Clock, and Mi Smart Clock. Specifically, we detect a \mic-like clock frequency, in \textit{both} the \textit{muted} as well as the \textit{unmuted} phases (see the \fclk~ column in Table~\ref{tbl:eval-other-devices}). In these devices, it is likely that the mute functionality is implemented in software, and hence the clock signal remains uninterrupted even during the mute phase. We believe that such an implementation is possible in devices such as smart speakers that do not have any power constraints (as they are always plugged-in to a power source).}

\begin{table}[]
\begin{tabular}{|l|c|c|c|c|}
\hline
SDR &
  RTL-SDR &
  \begin{tabular}[c]{@{}c@{}}SDRPlay \\ RSP-1A\end{tabular} &
  \begin{tabular}[c]{@{}c@{}}AirSpy HF+ \\ Discovery\end{tabular} &
  \begin{tabular}[c]{@{}c@{}}USRP \\ B210\end{tabular} \\ \hline
\begin{tabular}[c]{@{}l@{}}Cost\\ (USD) \end{tabular}                                                               & 40  & 140 & 169 & 1800 \\ \hline
\begin{tabular}[c]{@{}l@{}}Freq \\ Range\\ in MHz\end{tabular} &
  0.5 - 28.8 &
  0.001 - 2000 &
  0.0005 - 31 &
  70 - 6000 \\ \hline
\begin{tabular}[c]{@{}l@{}}Maximum\\ Bwidth\\ in MHz\end{tabular}      & 2.4    & 10     & 0.66   & 56      \\ \hline
\begin{tabular}[c]{@{}l@{}}Sampling \\ Mode\end{tabular}                  & Direct & I/Q    & I/Q    & I/Q     \\ \hline
\begin{tabular}[c]{@{}l@{}}Captured\\ Bwidth\\ in MHz\end{tabular}  & 14.4   & 30     & 23.76  & 30      \\ \hline
\begin{tabular}[c]{@{}l@{}}No. of \\ EM Traces\\ (in 3 mins)\end{tabular} & 490    & 650    & 115    & 2600    \\ \hline
TPR (\%)                                                                  & 67.6   & 99.7   & 98.8   & 94.1    \\ \hline
\begin{tabular}[c]{@{}l@{}}Form \\ Factor\end{tabular}                  & \begin{tabular}[c]{@{}l@{}}69 x 27\\x 13$mm^3$\end{tabular} & \begin{tabular}[c]{@{}l@{}}95 x 80\\x 30$mm^3$ \end{tabular}   & \begin{tabular}[c]{@{}l@{}}45 x 60\\x 10$mm^3$\end{tabular}    & \begin{tabular}[c]{@{}l@{}}9.7 x 15.5\\x 1.5$cm^3$\end{tabular}     \\ \hline
\end{tabular}
\caption{Table depicts the features of four popular software-defined radios that we utilize to validate the generalizability of \name across different signal-capture hardware.}
\label{tbl:eval-sdr}
\end{table}

%
\section{Performance due to Different Software Defined Radios}
\label{app:eval-sdr}
To verify \name's generalizability across signal-capture hardware, we evaluate its performance on four different software-defined radios (SDRs), namely, RTL-SDR, SDRPlay RSP1A, AirSpy HF+, and USRP B210 (with NooElec upconverter)~\cite{nooelec, usrp, sdrplay, airspy, rtlsdr}. In Table~\ref{tbl:eval-sdr}, we enlist their features, i.e., their cost, supported frequency range, maximum available bandwidth (in a single sweep), as well as their sampling mode (direct vs I/Q). 

As different SDRs may have different signal levels, we perform a short calibration for each SDR 
in order to compute a suitable amplitude threshold ($\theta_a$), for detecting the microphone clock frequency,~\fclk. 
Subsequently, we collect EM traces for three minutes from each SDR to report the obtained {\em True Positive Rate}. From the table, we observe that except for RTL-SDR, we obtain a {\em TPR} over 94\% for all other SDRs depicting the generalizability of \name across different hardware. 
In RTL-SDR, as they utilize direct-sampling, {\em aliasing} occurs, which leads to false detection of additional clock frequencies, thereby lowering the overall {\em TPR} to 67.6\%.  

\end{document}